\documentclass[a4paper,fleqn,usenatbib]{mnras}
\usepackage[T1]{fontenc}
\usepackage{ae,aecompl}
\usepackage{graphicx}
\usepackage{amsmath}
\usepackage{amssymb}

\title[Study of planetary nebulae MaC 2-1 and Sp 4-1]{Compact planetary nebulae MaC 2-1 and Sp 4-1: Photoionization models and dust characteristics}

\author[Rahul Bandyopadhyay et al.]{
Rahul Bandyopadhyay,\thanks{E-mail: rahul@bose.res.in}
Ramkrishna Das and
Soumen Mondal
\\
S. N. Bose National Centre for Basic Sciences, Block JD, Sector III, Salt Lake, Kolkata 700106, India\\
}

\date{Accepted XXX. Received YYY; in original form ZZZ}

\pubyear{0000}

\begin{document}
\label{firstpage}
\pagerange{\pageref{firstpage}--\pageref{lastpage}}
\maketitle

\begin{abstract}
We study the characteristics of planetary nebulae (PNe), MaC 2-1 and Sp 4-1. We use our optical spectra taken at 2 m Himalayan Chandra Telescope, Spitzer mid-infrared (mid-IR) spectra, HST images, and IR photometric data. These PNe have not been individually studied in details earlier. Both the PNe are in the low- to moderate-excitation class. MaC 2-1 shows the presence of silicon carbide (SiC) and magnesium sulphide (MgS) dust. Sp 4-1 hosts polycyclic aromatic hydrocarbon (PAH) molecules. We obtain plasma properties of the PNe from the optical and mid-IR emission line fluxes. We compute photoionization models of the PNe for self-consistent estimation of physical parameters associated with the central star and the nebula, including nebular abundances. From the modelling of the IR data, we obtain the characteristics of dust and molecules formed in the nebulae. From our study, we estimate that the progenitors of MaC 2-1 and Sp 4-1 had masses of 1.2 and 1.55 $M_{\sun}$, respectively, and both of them seem to have born in metal poor environment. Both are distant PNe, with the estimated distances of 16 and 18 kpc for MaC 2-1 and Sp 4-1, respectively.       
\end{abstract}


\section{Introduction}
Planetary nebulae (PNe) are formed out of the expelled outer layers of evolved stars, whose progenitors had 
main sequence masses $\sim1-8$ $M_{\sun}$. Study of PNe as a whole gives significant understanding on galactic chemical evolution through the knowledge about their progenitors up to birth-environment conditions and through prediction of enrichment of the surrounding interstellar medium (ISM) (e.g., \citealt{2006ApJ...651..898S}; \citealt{2016MNRAS.461..542G}). For this, having accurate values of physical parameters associated with the system is crucial. For an individual PN, these quantities are derived from correlated study of central star and nebula, which often include direct analysis of the observational data as well as modelling to reproduce the observables and self-consistent estimate of parameters. Also, a multiwavelength perspective is necessary to obtain even more complete understanding of a PN due to the pre-dominance of certain properties in particular wavelength intervals. For example, the main ionization structure of the nebula can be well mapped in the optical, while the dust and molecular features are pre-dominant in the infrared (IR) region of the spectrum. Few examples of PNe, for which such studies have been done are, IC 418 (\citealt{2009A&A...507.1517M} and \citealt{2018A&A...617A..85G}); NGC 6781 \citep{2017ApJS..231...22O}; SaSt 2-3 \citep{2019MNRAS.482.2354O}. However, the number of such PNe is small compared to the large number of PNe ($\sim$3500) discovered so far through various surveys and follow up diagnostics (e.g., \citealt{1992secg.book.....A}; \citealt{2001A&A...378..843K}; \citealt{2006MNRAS.373...79P}; \citealt{2008MNRAS.384..525M}; \citealt{2005MNRAS.362..753D}). The majority of PNe have not been investigated, mostly due to their faintness in general.   

In order to study the fainter and less-studied PNe, we have taken up an observational programme using 2 m Himalayan Chandra Telescope (HCT), Indian Astronomical Observatory (IAO, operated by Indian Institute of Astrophysics), Hanle, India. The programme 
This paper is the continuation of the work on the detailed study of the PNe \citep{10.1093/mnras/staa1518}, which were not studied in details earlier. For the present study, we have selected two PNe, MaC 2-1 (PN G$205.8-26.7$) and Sp 4-1 (PN G$068.7+14.8$). We study these PNe in detail with the help of our own observations and available archival data sets (Sec. \ref{sec:obsdataset}).
Both the PNe are compact (\citealt{2016ApJ...830...33S}, hereafter SSV16), depict apparently simple and round dual shell morphology, and appear to be consisting of a denser inner shell surrounded by a rarer outer shell. The selection of these two PNe is also based upon the availability of Spitzer spectra, and presence of the dust features: amorphous dust in both the PNe; silicon carbide (SiC) and magnesium sulphide (MgS) in MaC 2-1; polycyclic aromatic hydrocarbon (PAH) in Sp 4-1. This gives us scope to study and model the most common dust features seen among PNe. For each PN, we construct a physical model that closely replicates the characteristic observables of the PN, and hence, describe the object in a compact and self-consistent way. We model the PNe using the versatile and efficient photoionization code CLOUDY (version 17.00, \citealt{2017RMxAA..53..385F}, and references therein), where one can produce significant result assuming spherical geometry. The photoionization models are constrained by parameters of the simplistic nebular models obtained using the 3D modelling code SHAPE \citep{2011ITVCG..17..454S}. The results obtained from the individual studies of MaC 2-1 and Sp 4-1 are described in detail in Sections \ref{sec:mac2-1} and \ref{sec:sp4-1}, respectively. We discuss our results in Sec. \ref{sec:discussion} and finally, summarize our main findings in Sec. \ref{sec:summary}.

\begin{table}
\centering
\small
\caption{Log of HCT Observations. \label{tab:logobs}}
\begin{tabular}{l c c c c}
\hline
Object & Grism & Exposure (s) & Observation date\\
\hline
MaC 2-1 & Gr. 7 & 1500 & Dec. 26, 2019\\
& Gr. 8 & 1500 & Dec. 26, 2019\\
& Gr. 7 & 1500 & Oct. 8, 2020\\
& Gr. 8 & 1500 & Oct. 8, 2020\\
Sp 4-1 & Gr. 7 & 2100 & Oct. 7, 2020\\
& Gr. 8 & 2100 & Oct. 7, 2020\\
& Gr. 7 & 210 & Oct. 7, 2020\\
& Gr. 8 & 210 & Oct. 7, 2020\\
\hline
\end{tabular}
\end{table}

\section{Observational data set} \label{sec:obsdataset}

\subsection{2 m HCT optical spectra} \label{sec:hctspec}
Optical spectroscopic observations were carried out using the Hanle Faint Object Spectrograph Camera (HFOSC) instrument installed at 2 m HCT. Long-slit spectra were obtained in two spectral regions using the grisms Gr. 7, covering $\sim$3700-7000 {\AA} with resolution, $R\sim1400$ and Gr. 8, covering $\sim$5500-9000 {\AA} with $R\sim2200$. The log of observations is given in Table \ref{tab:logobs}. The slit was $1^{\prime\prime}.92$ in width and $\times 11^{\prime}$ in length, and was placed through the central stars of the PNe during the observations. The spectra of MaC 2-1 were obtained by taking two 1500 s exposure frames per grism. For Sp 4-1, one frame per grism was taken with 2100 s exposure, where the strongest lines were saturated. Hence, we took another set of observation with 210 s exposure to trace the strongest lines and use the longer exposure frames for the fainter lines. FeAr and FeNe lamp spectra were taken for the wavelength calibrations to the Gr. 7 and Gr. 8 spectra, respectively. Optical spectroscopic standards, Feige 34 ($V=11.18$, Spectral type: DO) and Feige 110 ($V=11.8$, Spectral type: DOp) were observed for spectral flux calibration of MaC 2-1 and Sp 4-1, respectively.

Spectra were reduced using the standard routines under the IRAF package. Initial corrections to the data were done through bias subtraction and cosmic-ray removal. One dimensional spectra were extracted optimally using the task APALL, through aperture selection and background subtraction and tracing. The spectra were corrected for wavelength using the reference lamp spectral data. Flux calibration to the spectra were done by calculation of airmass at the time of observations and using standard flux data of the standard stars.  

The spectra are corrected for interstellar extinction using the relation
\begin{equation}
I(\lambda)=F(\lambda)10^{[c(\mathrm{H}\beta)f(\lambda)]}
\end{equation}
where, $I(\lambda)$ and $F(\lambda)$ denotes the intrinsic and observed fluxes, respectively. $c(\mathrm{H}\beta)$ is the logarithmic extinction at H$\beta$. $f(\lambda)$ presents the extinction function given by \cite{1989ApJ...345..245C}. The value of $c(\mathrm{H}\beta)$ is obtained in order to attain the theoretical Balmer line ratios in the dereddened spectrum. We adopt the theoretical intrinsic flux ratios, $I(\mathrm{H}\alpha)/I(\mathrm{H}\beta)=2.847$, $I(\mathrm{H}\gamma)/I(\mathrm{H}\beta)=0.469$, $I(\mathrm{H}\delta)/I(\mathrm{H}\beta)=0.469$, where the values correspond to the electron temperature, $T_\mathrm{e}=10^4$ K and electron density, $N_\mathrm{e}=10^4$ cm$^{-3}$ \citep{2006agna.book.....O}. We denote the value of $c(\mathrm{H}\beta)$ obtained using H$\alpha$, H$\gamma$, and H$\delta$ as, $c(\mathrm{H}\beta)_{\mathrm{H}\alpha}$, $c(\mathrm{H}\beta)_{\mathrm{H}\gamma}$, and $c(\mathrm{H}\beta)_{\mathrm{H}\delta}$, respectively. The final value of $c(\mathrm{H}\beta)$ is obtained as, 
\begin{equation}
c(\mathrm{H}\beta)=\frac{{c(\mathrm{H}\beta)_{\mathrm{H}\alpha}}{\frac{F(\mathrm{H}\alpha)}{F(\mathrm{H}\beta)}} + {c(\mathrm{H}\beta)_{\mathrm{H}\gamma}}{\frac{F(\mathrm{H}\gamma)}{F(\mathrm{H}\beta)}} + {c(\mathrm{H}\beta)_{\mathrm{H}\delta}}{\frac{F(\mathrm{H}\delta)}{F(\mathrm{H}\beta}}}{\frac{F(\mathrm{H}\alpha)}{F(\mathrm{H}\beta)} + \frac{F(\mathrm{H}\gamma)}{F(\mathrm{H}\beta)} + \frac{F(\mathrm{H}\delta)}{F(\mathrm{H}\beta}}
\end{equation}            

\subsection{Spitzer mid-IR spectra} \label{sec:spitzer}

From the Spitzer Heritage Archive (SHA)\footnote{\url{https://sha.ipac.caltech.edu/}}, we obtained calibrated mid-IR spectral data observed using Infrared Spectrograph (IRS; Houck et al. 2004) onboard Spitzer Space Telescope. The observations were made using the short-low (SL) module covering 5.2-14.5 $\mu$m (slit dimension $3^{\prime\prime}.6\times57^{\prime\prime}$) and long-low (LL) module covering 14.0-38.0 $\mu$m (slit dimension $10^{\prime\prime}.5\times168^{\prime\prime}$) (AORKEY: 25849600; Program Id: 50261; PI: Stanghellini, Letizia). 

\subsection{HST narrow band images} \label{sec:hstimg}
For our work, we use high-resolution Hubble Space Telescope (HST) images of the PNe from the Hubble Legacy Archive (HLA)\footnote{\url{https://hla.stsci.edu/}} (PI: Stanghellini; Proposal ID: 11657, SSV16). The images were obtained using Wide-field Camera 3 (WFC3) (Kimble et al. 2008) through F502N ($\lambda_p=5010\AA, \Delta\lambda=65\AA$) filter, which maps the region of [O~{\sc iii}] emission within the nebulae. The exposure time for each image was 60 seconds. MaC 2-1 was observed on Nov. 11, 2009 and Sp 4-1 was observed on Jul. 30, 2009.

\subsection{Photometric data} \label{sec:photodata}
We have collected photometric fluxes of the objects from AKARI, 2MASS, WISE point source catalogues available at Infrared Science Archive (IRSA)\footnote{\url{https://irsa.ipac.caltech.edu/}}. The observed photometric fluxes up to 3.3 $\mu$m are corrected for interstellar extinction using Equation 1. Beyond that, the extinction is considered negligible, particularly due to the small value of extinction coefficient for both the PNe.

For the purpose of our analysis, the data set are combined together in absolute flux scale. We scale out optical spectrum to the total H$\alpha$ fluxes of the PNe given by \citet{2013MNRAS.431....2F}. The flux scale of the Spitzer spectrum is well matched with the photometric flux values in that region. Hence, no further scaling is done.  

\section{Results: Study of MaC 2-1} \label{sec:mac2-1}
MaC 2-1 ($R.A.=05^h03^m41.88^s$, $Dec.=-06^{\circ}10^{\prime}03^{\prime\prime}.16$) is a PN almost circular in appearance with a radius of about $1.3^{\prime\prime}$. SSV16 have placed this PN in a morphological classification of `elliptical', and noticed the presence of microstructures, such as a bright interior ring, and ansae.  

\subsection{Spectral analyses}

\begin{figure}
\centering
\scalebox{0.65}[0.65]{\includegraphics{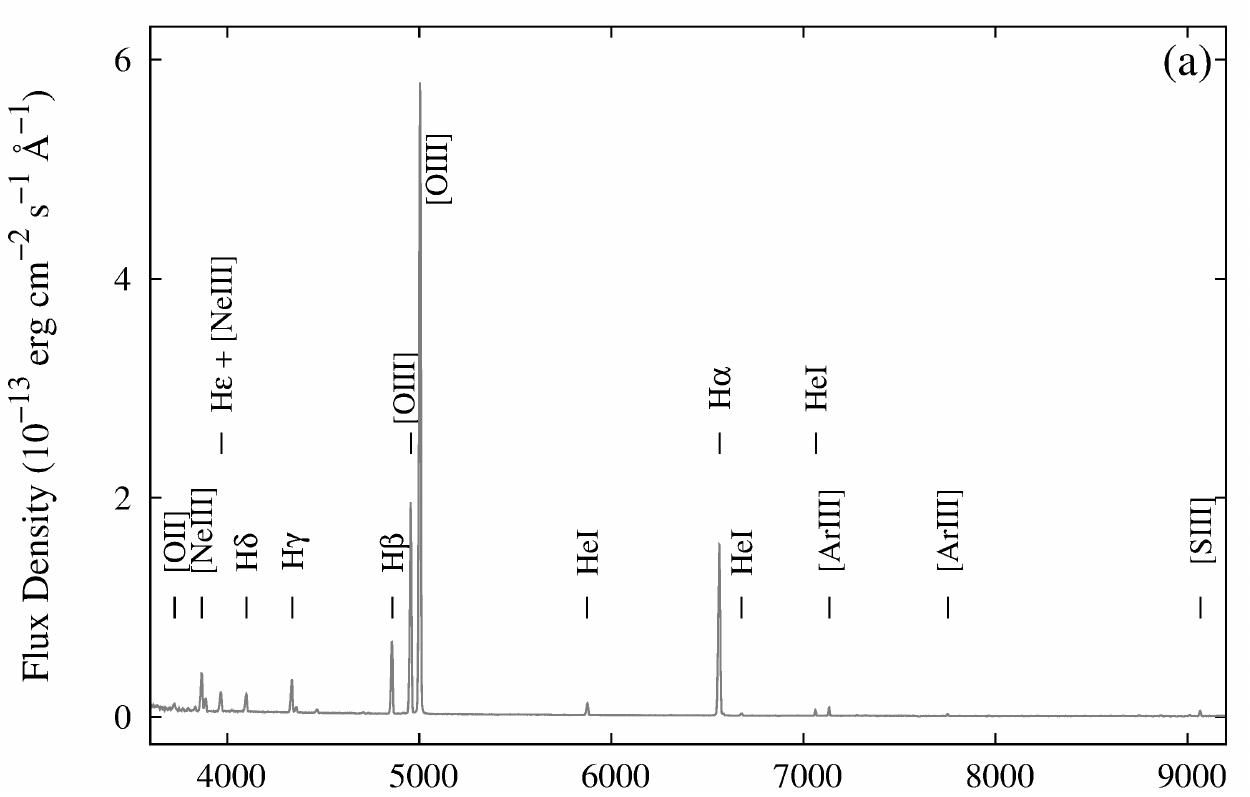}}

\scalebox{0.65}[0.65]{\includegraphics{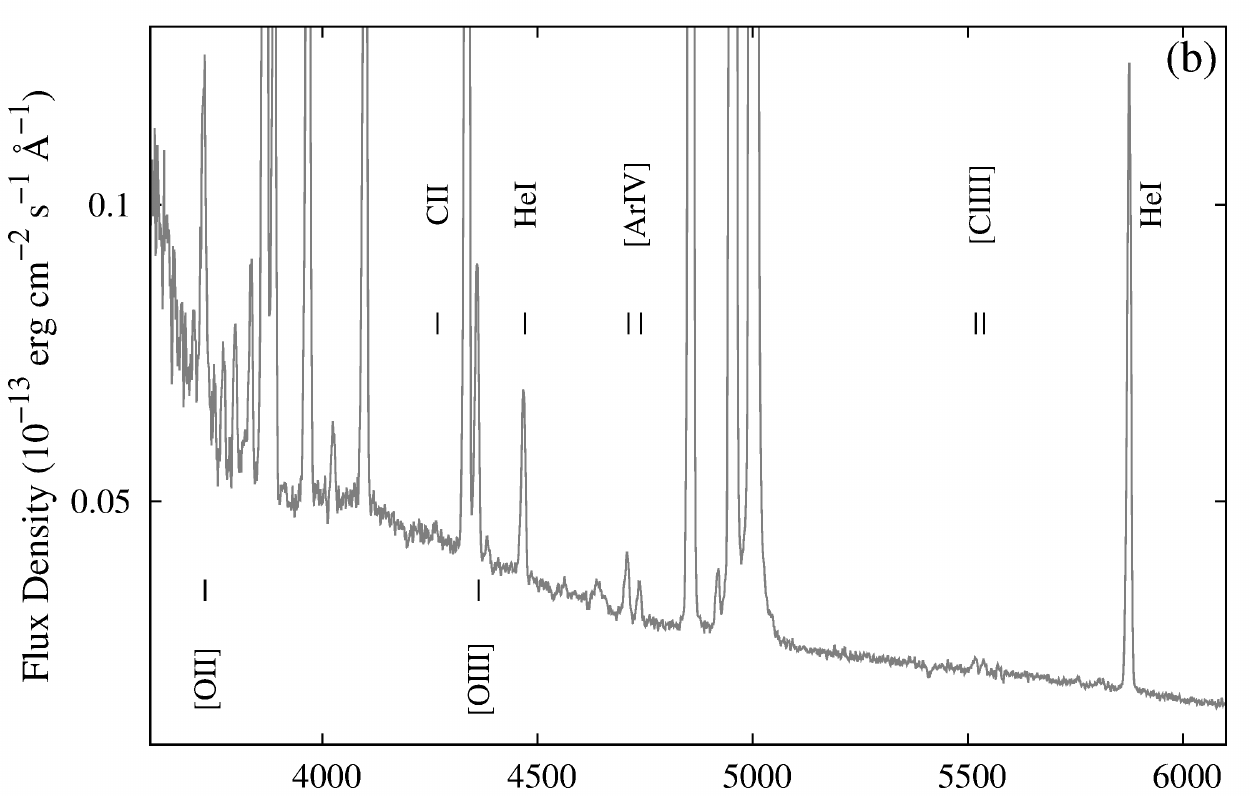}}

\scalebox{0.65}[0.65]{\includegraphics{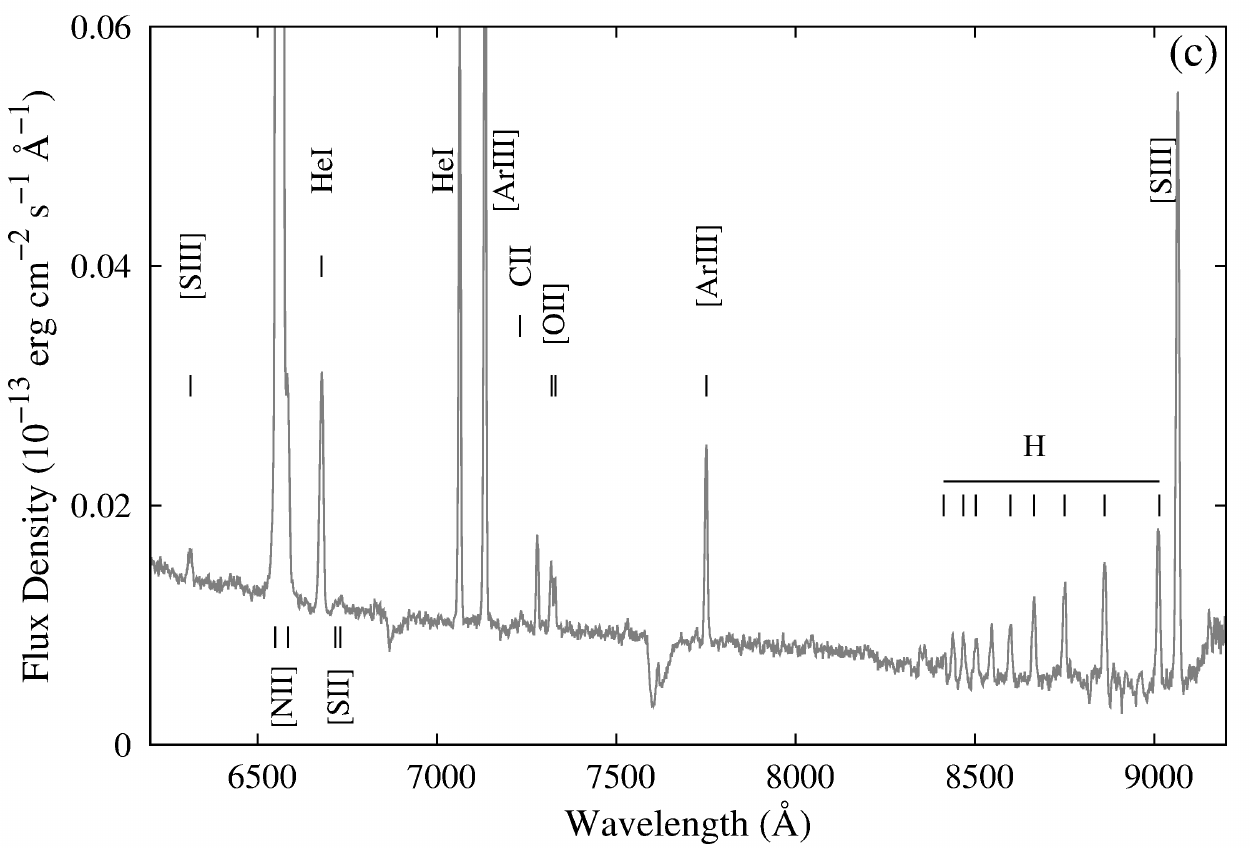}}
 \caption{The optical spectrum of MaC 2-1 in intrinsic absolute fluxes. Panel (a) shows the entire spectrum with strongest emission lines. The panels (b) and (c) show the magnified spectrum to emphasize the weaker emission lines in the blueward and redward regions, respectively. \label{fig:specoptmac2-1}}
\end{figure} 

\subsubsection{Interstellar extinction coefficient}
The effects of interstellar extinction on MaC 2-1 is found out to be quite small. We find $c(\mathrm{H}\beta)_{\mathrm{H}\alpha}=0.11$, $c(\mathrm{H}\beta)_{\mathrm{H}\gamma}=0.11$, and $c(\mathrm{H}\beta)_{\mathrm{H}\delta}=0.15$ using Equation 1, and thus, obtain a value of $c(\mathrm{H}\beta)\simeq0.11$, using Equation 2. Earlier, \citet{1996A&A...307..215C} (hereafter C96) and \citet{2013MNRAS.431....2F} adopted the $c(\mathrm{H}\beta)=0.0$. Hence, we find no proper estimation of $c(\mathrm{H}\beta)$ for MaC 2-1 from previous studies. However, \citet{2016MNRAS.455.1459F} (hereafter FPB16) reported the extinction $E(B-V)=0.08\pm0.29$, which corresponds to a $c(\mathrm{H}\beta)\sim0.1$, very close to our estimated value.    

\subsubsection{Spectral features} \label{sec:specfeatures}
{\textit{Optical spectrum}:} The optical spectra (Fig. \ref{fig:specoptmac2-1}) of MaC 2-1 features a low- to moderate-excitation nebula. The detected line list in the spectrum include the commonly observed strong recombination lines (RLs), such as hydrogen Balmer lines, H$\alpha$ 6563, H$\beta$ 4861, H$\gamma$ 4340, H$\delta$ 4101, H$\epsilon$ 3970 {\AA}; weaker Paschen lines within in the range of $8300-9100$ {\AA}, and helium RLs, He~{\sc i} 4471, 5876, 6678, 7065 {\AA}. We observe weak presence of C~{\sc ii} 4267, 7231 {\AA}. The central star temperature seems to be on the lower side, presumably $\lesssim55,000$ K, as He~{\sc ii} emission is not detected in our spectrum. From the flux of the stellar continuum and $\mathrm{H}\beta$, \cite{1989A&A...222..237G} reported a Zanstra temperature of $T_\mathrm{Z}=33000$ K for the central star of MaC 2-1. They also put the PN in excitation class (EC) of 5 according to the well-known classification by \citet{1956gn.book.....A} (hereafter A56), which however, implies the detection of He~{\sc ii} emission. Using our flux values of [O~{\sc iii}] 4959, 5007 and $\mathrm{H}\beta$, we have $EC=3$ according to A56. We also obtain the excitation class as $EC=3.91$ using the expression derived by \citet{1990ApJ...357..140D} from the scheme given by \citet{1984MNRAS.208..633M} (hereafter M84). Hence, the central star might be at a critical temperature, just been able to form a very weak He~{\sc ii}, undetected in our spectrum. Among the collisionally excited lines (CELs), we prominently observe [N~{\sc ii}] 6548, 6583 {\AA}; [O~{\sc iii}] 5007, 4959 {\AA} and the auroral [O~{\sc iii}] 4363 {\AA}; [Ne~{\sc iii}] 3869 {\AA}; [S~{\sc iii}] 6312, 9069 {\AA}; [Cl~{\sc iv}] 7531, 8046 {\AA}; [Ar~{\sc iii}] 7136, 7751 {\AA}; and the doublets [S~{\sc ii}] 6716, 6731 {\AA}; [Cl~{\sc iii}] 5518, 5538 {\AA}, and [Ar~{\sc iv}] 4711, 4740 {\AA}. The [N~{\sc ii}] lines are partially blended with H$\alpha$. The auroral [N~{\sc ii}] 5755 {\AA} is not detected with confidence in our spectrum. 

\begin{figure}
\centering
\scalebox{0.65}[0.65]{\includegraphics{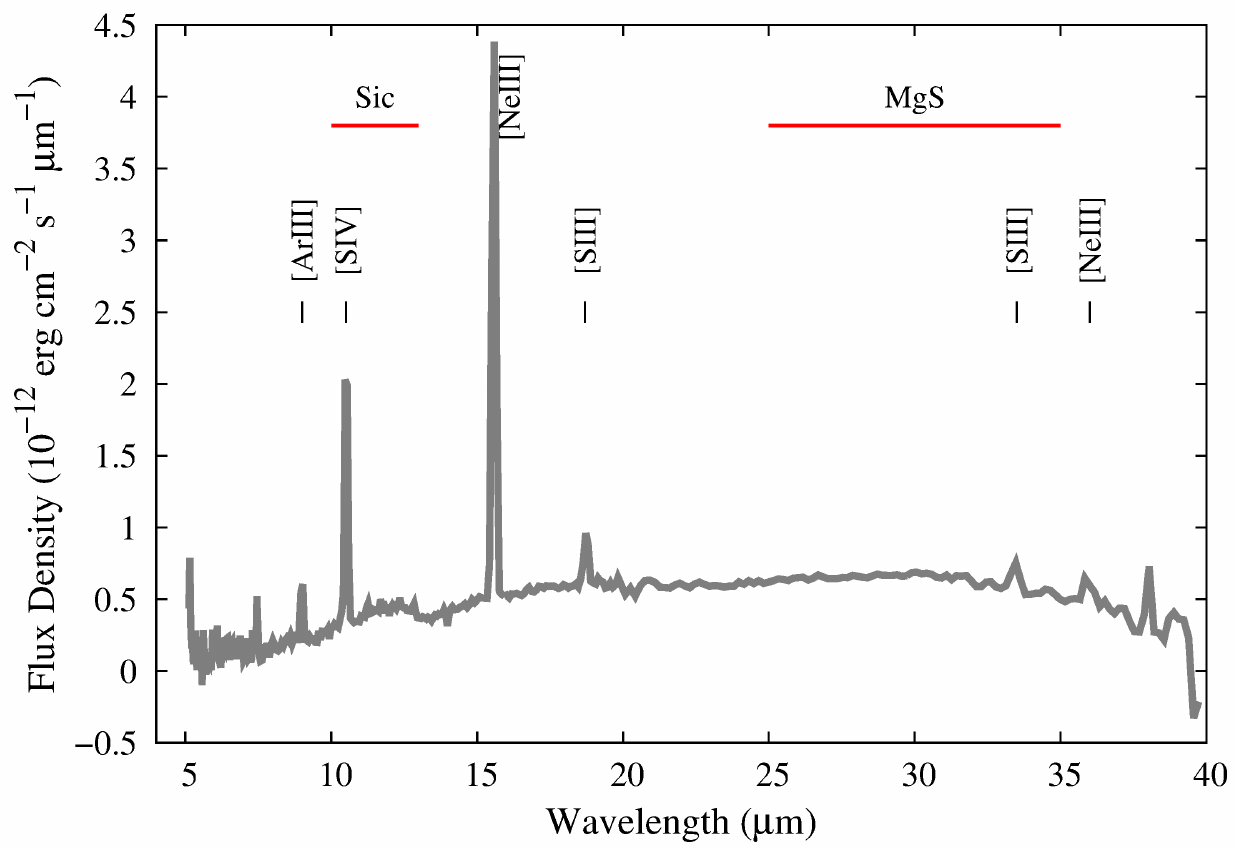}}
 \caption{The mid-IR Spitzer spectrum of MaC 2-1. Prominent features including the broad dust features are marked. Fluxes are in absolute scale. \label{fig:specmirmac2-1}}
\end{figure} 

{\textit{IR spectrum}:} The Spitzer spectrum (Fig. \ref{fig:specmirmac2-1}) shows emission lines from [Ne~{\sc iii}] 15.6, 36.0 $\mu$m, [Ar~{\sc iii}] 9.0 $\mu$m, [S~{\sc iii}] 18.7, 33.5 $\mu$m, [S~{\sc iv}] 10.5 $\mu$m. The mid-IR rise of the continuum evident in the spectrum is generally due to thermal emission carbonaceous dust present within the nebula. In a characterization scheme of PNe according to their dust nature observed in the Spitzer spectra, \citet{2012ApJ...753..172S} (hereafter S12) classified MaC 2-1 into the category having carbon rich dust of aliphatic nature. Broad band features are also observed in the spectrum as well. The broad 11.5 $\mu$m feature, generally attributed to SiC, is weakly present in the spectrum. The shape of spectrum in the $23-38$ $\mu$m region suggests the presence of the broad 30 $\mu$m MgS feature. From, our photoionization modelling (Sec. \ref{sec:pimodelmac2-1}), we satisfactorily reproduce both the 11.5 and 30 $\mu$m features considering SiC and MgS grains, respectively.

\begin{figure}
\scalebox{0.5}[0.5]{\includegraphics{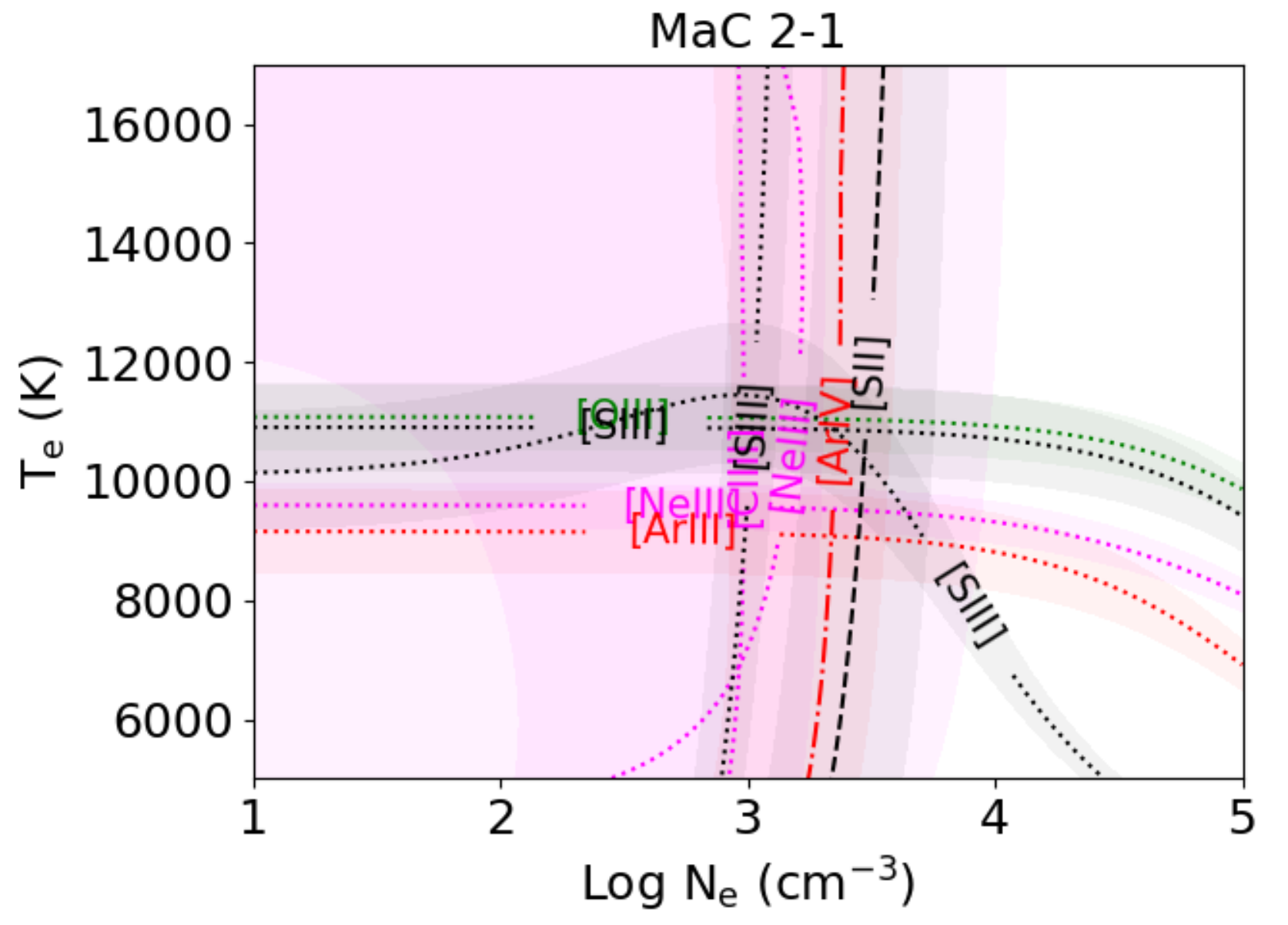}}
 \caption{The $T_\mathrm{e}$ vs Log $N_\mathrm{e}$ plot for MaC 2-1. The labels with the graphs denote the corresponding ion used for generating the same. \label{fig:tenemac2-1}}
\end{figure}

\begin{table}
\centering
\small
\caption{Temperatures and Densities for MaC 2-1. \label{tab:tenemac2-1}}
 \begin{tabular}{lccc}
 \hline
& Diagnostic line id. & Ratio & $T_\mathrm{e}$/$N_\mathrm{e}$ value\\
 \hline
$T_\mathrm{e}$ & [O~{\sc iii}] 4363/5007 & 0.0088 & ${10970}$\\
$T_\mathrm{e}$ & [S~{\sc iii}] 6312/9069 & 0.081 & ${10890}$\\
$T_\mathrm{e}$ & [Ne~{\sc iii}] 3869/15.6m & 0.599 & ${10000}$\\
$T_\mathrm{e}$ & [Ar~{\sc iii}] 7136/9.0m & 0.911 & ${10000}$\\
$T_\mathrm{e}$ & [S~{\sc iii}] 9069/18.7m & 0.833 & ${13050}$\\
 \hline
$N_\mathrm{e}$ & [Ar~{\sc iv}] 4740/4711 & 0.94 & ${2280}$\\
$N_\mathrm{e}$ & [Cl~{\sc iii}] 5538/5518 & 0.825 & ${990}$\\
$N_\mathrm{e}$ & [S~{\sc ii}] 6731/6716 & 1.66 & ${3210}$\\
$N_\mathrm{e}$ & [Ne~{\sc iii}] 15.6m/36.0m & 11.537 & ${1440}$\\
$N_\mathrm{e}$ & [S~{\sc iii}] 18.7m/33.5m & 1.286 & ${1110}$\\
 \hline
 \end{tabular}
\end{table} 

\begin{table}
\centering
\small
\caption{Nebular ionic and total abundances of MaC 2-1 from direct method ($A(B)=A\times10^B$). \label{tab:abunmac2-1}}
\begin{tabular}{l c c c c c}
\hline
X & X$^{+}$/H & X$^{2+}$/H & X$^{3+}$/H & ICF & X/H\\
\hline
He	&	0.108	&	-	 	&	-		& 1.0 	& 0.11\\
C	&	-		&	-		&	-	    & - 		& - \\
N	&	6.31(-7)	&	-		&	-		& 164 	& 1.03(-4)\\
O	&	5.79(-6)	&	2.20(-4)	&	-		& 1.0 	& 2.26(-4)\\
Ne	&	-		&	4.22(-5)	&	-		& 1.38 	& 5.83(-5)\\
S	&	8.59(-9)	&	9.01(-7)&	1.13(-6)	& 2.17 		& 1.98(-6)\\
Cl	&	-		&	3.69(-8)	&	1.16(-8)	& 2.26 		& 8.34(-8)\\
Ar	&	-		&	3.73(-7)&	2.48(-7)	& 1.67 	& 6.21(-7)\\
\hline
\end{tabular}
\end{table}

\begin{table}
\centering
\small
\caption{Model results for MaC 2-1. \label{tab:pimodelmac2-1}}
\begin{tabular}{l c c c c r}
\hline
Geometry &&&&& Spherical\\
$d$ (kpc) &&&&& 16.0\\
$T_\mathrm{eff}$ (K) &&&&& 49500\\
$L$ $(L_{\sun})$ &&&&& 3900\\
$R$ $(R_{\sun})$ &&&&& 0.85\\
$M$ $(M_{\sun})$ &&&&& $\sim$0.57\\
$M_\mathrm{pr}$ $(M_{\sun})$ &&&&& 1.20\\
Log $g$ $(\mathrm{cm}$ $\mathrm{s^{-2}})$ &&&&& 4.34\\
$r_\mathrm{in}$ $(\mathrm{pc})$ &&&&& 0.021\\
$r_\mathrm{inter}$ $(\mathrm{pc})$ &&&&& 0.035\\
$r_\mathrm{out}$ $(\mathrm{pc})$ &&&&& 0.116\\
$n_\mathrm{H}$ (at $r_\mathrm{in}$) $(10^3$ $\mathrm{cm^{-3}})$ &&&&& 4.0 \\
$f$ &&&&& 1.0\\
$m_\mathrm{cloud}$ $(M_{\sun})$ &&&&& 0.15\\
\hline
\hline
\end{tabular}
\begin{tabular}{l c c c c}
\multicolumn{5}{c}{Chemical composition}\\
\hline
& \multicolumn{3}{c}{Model ionic fractions} &\\
X & X$^{+}$ & X$^{2+}$ & X$^{3+}$ & X/H \\
\hline
He	&	1.0000	&	0.0001	&	-	&	0.12	\\
C	&	0.0104	&	0.9528	&	0.0370	&	8.07(-5)\\
N	&	0.0085	&	0.9419	&	0.0491	&	7.64(-5)	\\
O	&	0.0176	&	0.9817	&	0.0000  &	2.45(-4)	\\
Ne	&	0.0442	&	0.9550	&	0.0000	&	6.38(-5)	\\
S	&	0.0044	&	0.6109	&	0.3828	&	1.72(-6)	\\
Cl	&	0.0054	&	0.6950	&	0.2992	&	4.74(-8)	\\
Ar	&	0.0013	&	0.6761	&	0.3221	&	8.28(-7)\\
\end{tabular}
\begin{tabular}{l c c c c}
\hline
Dust types & $T_\mathrm{d}$ (K) & $D/G$ & X & (X/H)$_{dust}$\\
\hline
AC & 62-210 & 3.38(-5) & C & 1.27(-5)\\
Graphite & 52-205 & 6.77(-5) & O & 2.62(-6)\\
Silicate & 51-148 & 7.62(-5) & Mg & 1.66(-6)\\
SiC & 104-188 & 4.42(-6) & Si & 8.20(-7)\\
MgS & 67-174 & 4.49(-5) & S & 1.12(-6)\\
All dust & 51-210 & 2.27(-4) & Fe & 7.68(-7) \\
\hline
\end{tabular}
\end{table} 

\begin{table}
\centering
\small
\caption{Comparison between observed and modeled line fluxes of MaC 2-1. Observed and modelled flux values are given with respect to $I(\mathrm{H}\beta)=100$. \label{tab:obsvsmodmac2-1}}
\begin{tabular}{lccc}
\hline
Observables & Observed & Modelled & $\kappa_\mathrm{i}$\\
& flux & flux & \\
\hline
Log $I(\mathrm{H}\beta)$ & -12.15 & -12.09 & \\
	H$\delta$			4101	{\AA}	&	25.46	&	26.45	&	-0.40	\\
	H$\gamma$			4340	{\AA}	&	47.51	&	47.38	&	0.03	\\
	H$\beta$			4861	{\AA}	&	100.00	&	100.00	&	0.00	\\
	H$\alpha$			6563	{\AA}	&	282.91	&	276.87	&	0.23	\\
$\mathrm{	He~{\sc	I}	}$	4026	{\AA}	&	2.61	&	2.79	&	-0.36	\\
$\mathrm{	He~{\sc	I}	}$	4388	{\AA}	&	0.77	&	0.74	&	0.15	\\
$\mathrm{	He~{\sc	I}	}$	4471	{\AA}	&	5.00	&	5.90	&	-0.91	\\
$\mathrm{	He~{\sc	I}	}$	5876	{\AA}	&	16.31	&	16.36	&	-0.04	\\
$\mathrm{	He~{\sc	I}	}$	6678	{\AA}	&	3.65	&	4.27	&	-0.85	\\
$\mathrm{	He~{\sc	I}	}$	7065	{\AA}	&	3.86	&	9.93	&	-5.19	\\
$\mathrm{	He~{\sc	I}	}$	7281	{\AA}	&	0.63	&	0.99	&	-1.74	\\
$\mathrm{	[N~{\sc	II}]	}$	5755	{\AA}	&	-	&	0.06	&	-	\\
$\mathrm{	[N~{\sc	II}]	}$	6548	{\AA}	&	-	&	1.21	&	-	\\
$\mathrm{	[N~{\sc	II}]	}$	6583	{\AA}	&	3.85	&	3.57	&	0.41	\\
$\mathrm{	[O~{\sc	II}]	}$	3727	{\AA}	&	12.32	&	11.63	&	0.61	\\
$\mathrm{	[O~{\sc	II}]	}$	7325	{\AA}	&	1.04	&	0.80	&	1.42	\\
$\mathrm{	[O~{\sc	III}]	}$	4363	{\AA}	&	7.66	&	6.69	&	0.75	\\
$\mathrm{	[O~{\sc	III}]	}$	4959	{\AA}	&	286.83	&	280.14	&	0.25	\\
$\mathrm{	[O~{\sc	III}]	}$	5007	{\AA}	&	869.47	&	835.81	&	0.41	\\
$\mathrm{	[Ne~{\sc	III}]	}$	3869	{\AA}	&	56.53	&	56.88	&	-0.06	\\
$\mathrm{	[Ne~{\sc	III}]	}$	15.50	$\mu$m	&	94.44	&	71.44	&	2.93	\\
$\mathrm{	[Ne~{\sc	III}]	}$	36.00	$\mu$m	&	8.19	&	6.34	&	1.40	\\
$\mathrm{	[S~{\sc	II}]	}$	6716	{\AA}	&	0.11	&	0.10	&	0.42	\\
$\mathrm{	[S~{\sc	II}]	}$	6731	{\AA}	&	0.18	&	0.12	&	1.35	\\
$\mathrm{	[S~{\sc	III}]	}$	6312	{\AA}	&	0.61	&	0.53	&	0.51	\\
$\mathrm{	[S~{\sc	III}]	}$	9069	{\AA}	&	7.54	&	7.45	&	0.07	\\
$\mathrm{	[S~{\sc	III}]	}$	18.70	$\mu$m	&	9.06	&	7.97	&	0.70	\\
$\mathrm{	[S~{\sc	III}]	}$	33.50	$\mu$m	&	7.04	&	4.14	&	2.91	\\
$\mathrm{	[S~{\sc	IV}]	}$	10.50	$\mu$m	&	32.90	&	27.97	&	1.70	\\
$\mathrm{	[Cl~{\sc	III}]	}$	5518	{\AA}	&	0.45	&	0.29	&	1.67	\\
$\mathrm{	[Cl~{\sc	III}]	}$	5538	{\AA}	&	0.37	&	0.32	&	0.60	\\
$\mathrm{	[Cl~{\sc	IV}]	}$	7531	{\AA}	&	0.22	&	0.09	&	3.53	\\
$\mathrm{	[Cl~{\sc	IV}]	}$	8046	{\AA}	&	0.20	&	0.20	&	-0.13	\\
$\mathrm{	[Ar~{\sc	III}]	}$	7136	{\AA}	&	5.62	&	7.04	&	-1.24	\\
$\mathrm{	[Ar~{\sc	III}]	}$	7751	{\AA}	&	2.17	&	1.67	&	1.44	\\
$\mathrm{	[Ar~{\sc	III}]	}$	9	$\mu$m	&	6.17	&	5.67	&	0.46	\\
$\mathrm{	[Ar~{\sc	IV}]	}$	4711	{\AA}	&	1.93	&	1.27	&	2.30	\\
$\mathrm{	[Ar~{\sc	IV}]	}$	4740	{\AA}	&	1.13	&	0.96	&	0.87	\\
\hline
\end{tabular}
\end{table}     

\subsubsection{Electron temperatures, densities and CEL anbundances} \label{sec:pynebmac2-1} 
The package PyNeb \citep{2015A&A...573A..42L} is used to obtain few results from the analysis of emission line fluxes. Electron temperatures ($T_\mathrm{e}$) and electron densities ($N_\mathrm{e}$) are calculated using the known diagnostic emission line flux ratios. [O~{\sc iii}], [S~{\sc iii}], [Ne~{\sc iii}], and [Ar~{\sc iii}] are used for calculating $T_\mathrm{e}$ and [Ar~{\sc iv}], [Cl~{\sc iii}], [S~{\sc ii}], [Ne~{\sc iii}], and [S~{\sc iii}] are used for $N_\mathrm{e}$ calculations. The $T_\mathrm{e}$ and $N_\mathrm{e}$ values and corresponding line flux ratios that are used for the calculations are listed in Table \ref{tab:tenemac2-1}. Fig. \ref{fig:tenemac2-1} shows $T_\mathrm{e}$-$N_\mathrm{e}$ plot, which suggests a good solution to the values of these two nebular parameters. In this work, we entirely focus on deriving elemental abundances from CELs, and not from RLs. We use the values $T_\mathrm{e}\sim11,000$ K and $N_\mathrm{e}\sim2500$ cm$^{-3}$ for our calculation of abundances. We obtain the nebular line fluxes from different ionization states of elements present in the spectrum and calculate the ionic abundances of the elements (Table \ref{tab:abunmac2-1}). From the ionic abundances, we estimate the total abundances of the elements (Table \ref{tab:abunmac2-1}) by the method of using ionization correction factors (ICFs), and refer to it as `Direct Method' (as in B20) in the rest of this paper for discussion. We use the ICFs formulae given in \citet{2014MNRAS.440..536D}. We could not obtain C/H from direct method as there are no CEL of C in the spectral range we use in this work. 

\begin{figure}
\centering
\scalebox{0.24}[0.24]{\includegraphics{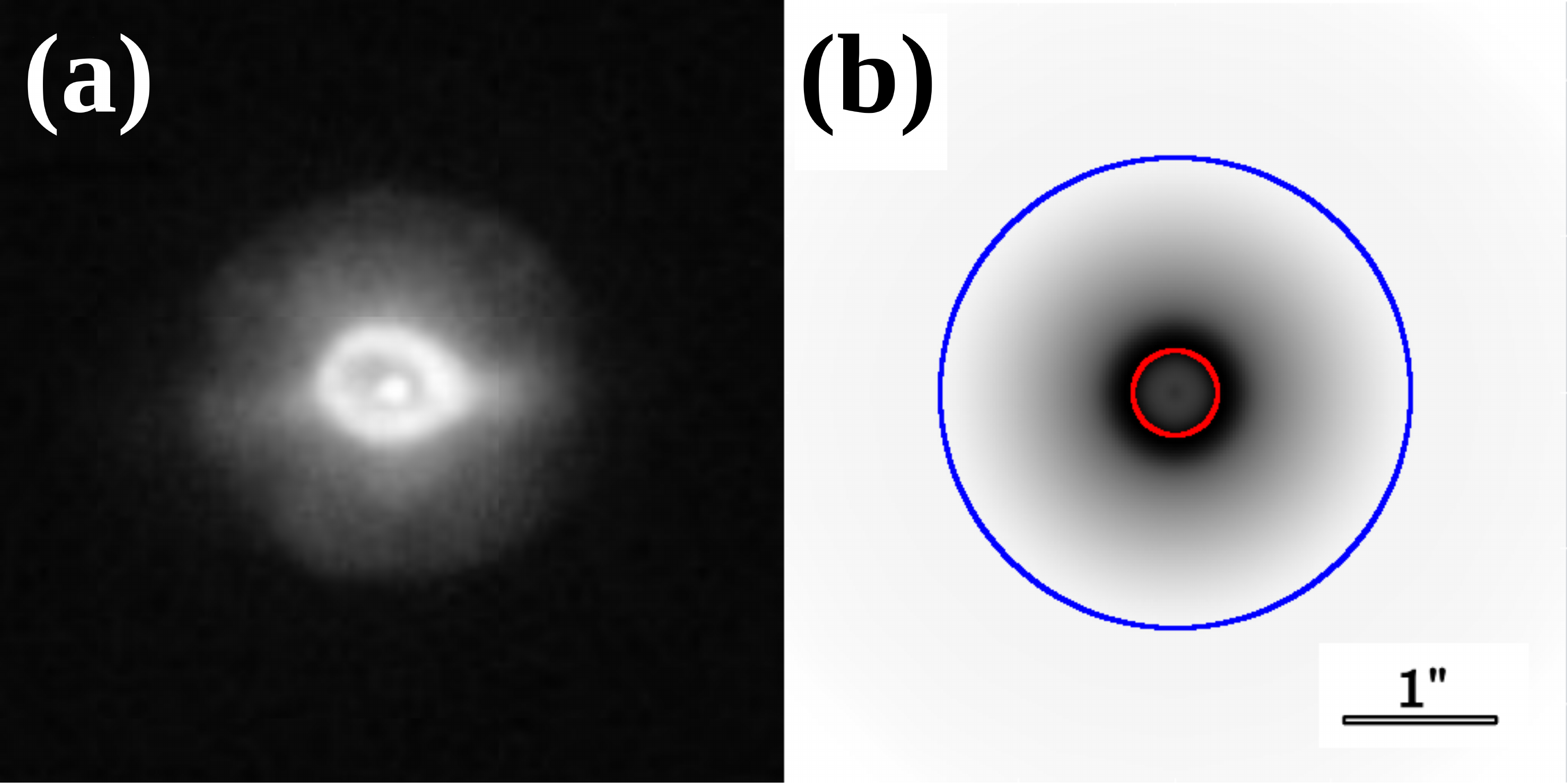}}

\scalebox{0.9}[0.9]{\includegraphics{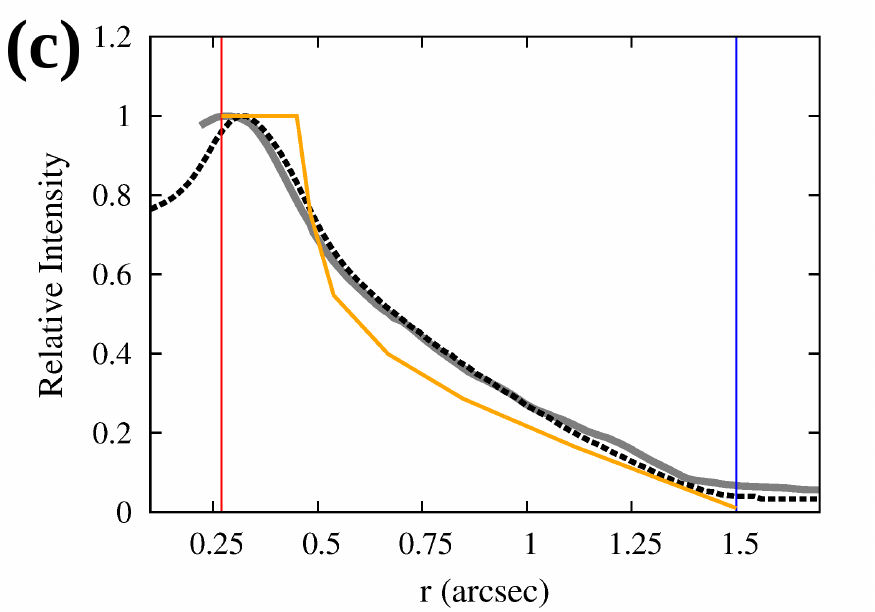}}
 \caption{(a) The HST image of MaC 2-1 in [O~{\sc iii}]. The image box measures $5^{\prime\prime}\times5^{\prime\prime}$. (b) The grey scale 3D modelled image of MaC 2-1 obtained using SHAPE. The concentric circles are the inner (in red) and outer (in blue) radii of the shell. (c) The grey solid line presents the observed radial intensity profile obtined from the HST image. The orange dotted line shows the radial density structure of the 3D model, that best-fits the observed radial intensity profile. The best-fitting modelled radial intensity profile (black dotted line) shows a good fit with the observed profile. The vertical red and blue lines denote the inner and outer radii, respectively, as mentioned above. In our CLOUDY modelling, the density of the shell is specified using the radial density struture obtained from the 3D model. Also, the $r_\mathrm{in}$ and $r_\mathrm{out}$ are calculated using the inner and outer radii of the 3D model (see details in Sec. \ref{sec:pimodelmac2-1}). \label{fig:denprofilemac2-1}}
\end{figure} 

\begin{figure*}
\centering
\scalebox{1.0}[1.0]{\includegraphics{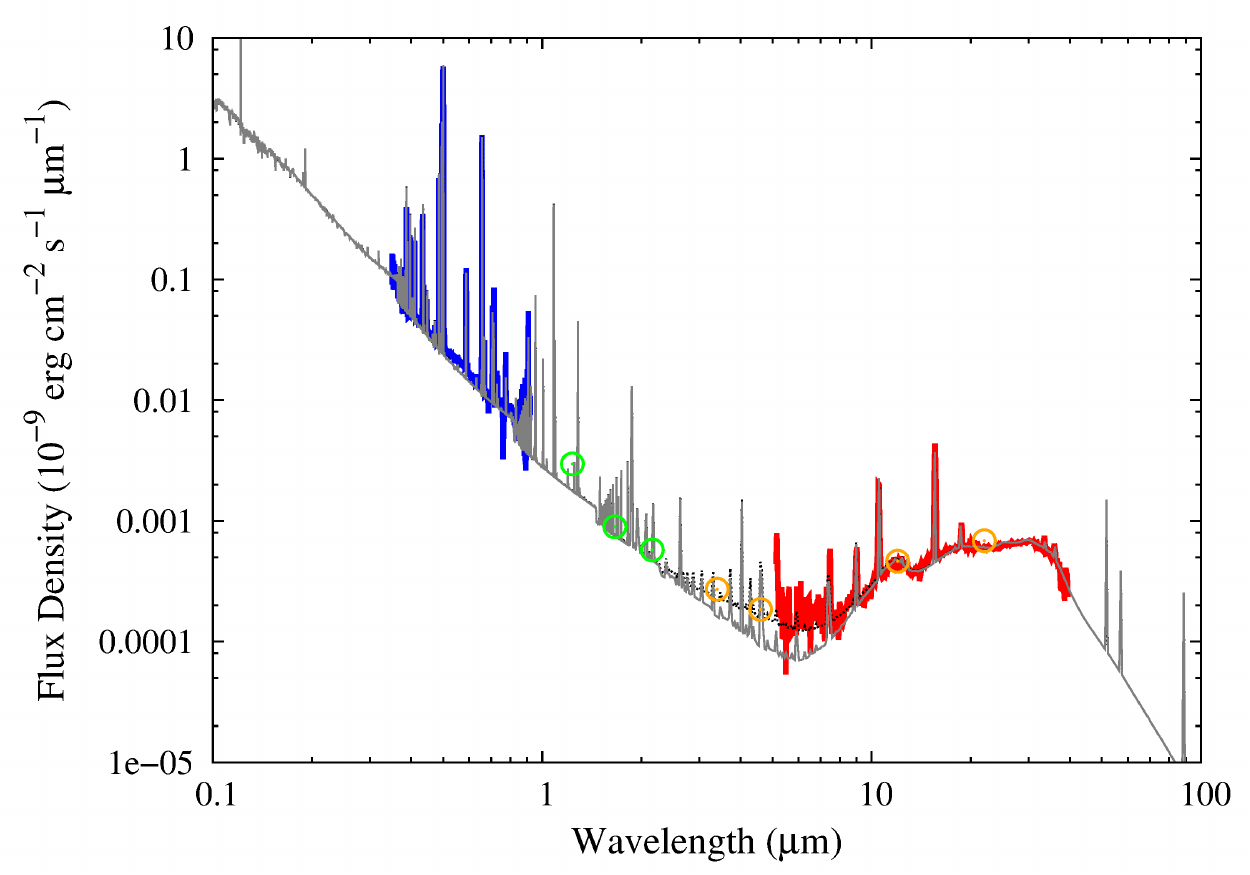}}
 \caption{The modelled spectra of MaC 2-1 in grey solid line is plotted with the observed data for comparison. The observed optical spectrum is shown in blue and mid-IR spectrum is shown in red. The green and orange circles denote the observed photometric data points from 2MASS and WISE point source catalogues, respectively. The black dotted line is the test model spectrum that includes warm dust region (see details in Sec. \ref{sec:testmodels}) \label{fig:pimodelmac2-1}}
\end{figure*} 

\begin{figure}
\centering
\scalebox{0.65}[0.65]{\includegraphics{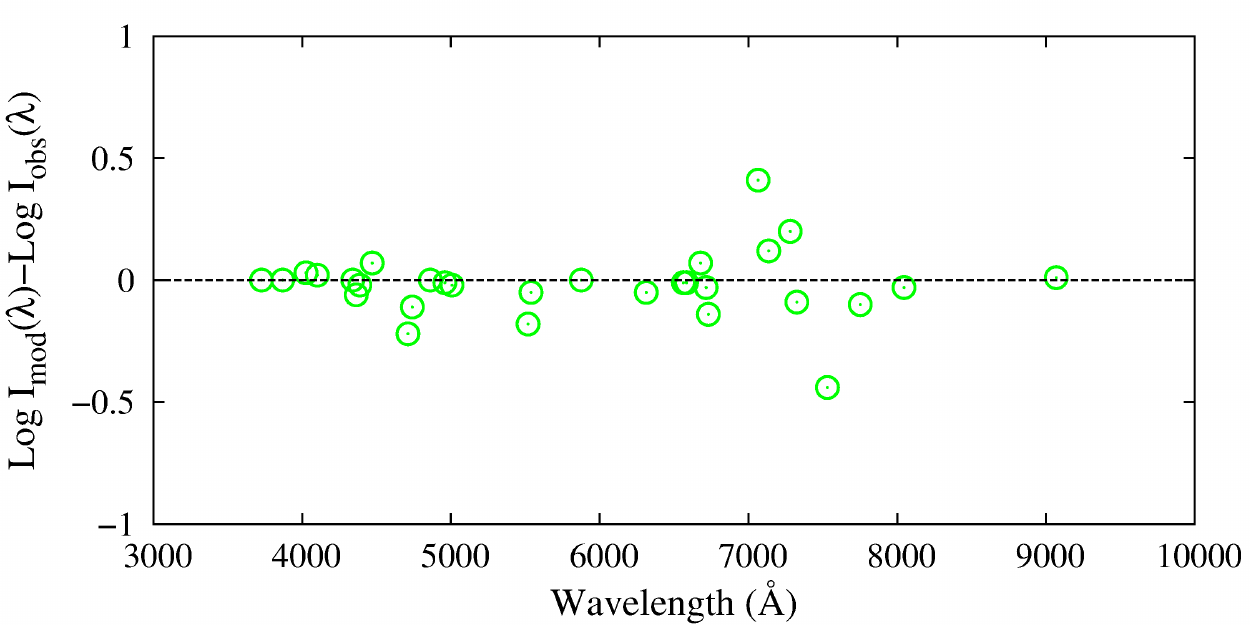}}
 \caption{The differences between logarithmic observed and modelled emission line flux are plotted along the wavelength. The points falling close to the zeroth line implies a good model fit. \label{fig:errormac2-1}}
\end{figure}

\subsection{Modelling of MaC 2-1} \label{sec:pimodelmac2-1}
We construct a physical model of MaC 2-1 that closely reproduces the observed physical quantities of this PN. We aim to obtain its characteristics in a compact and self-consistent way. We use the photoionization code CLOUDY for the photoionization modelling of the observed optical and mid-IR spectrum of MaC 2-1. 

For our CLOUDY models, we consider a spherical shell (nebula) around an ionizing source (central star). We specify a set of input physical parameters, which characterizes the central star and the nebula. CLOUDY internally solves the radiation transfer, ionization balance and energy balance equations at each point of the nebula, and calculate the model spectrum. A total model spectrum consists of the attenuated stellar spectrum of the central star, the nebular continuum and the emission lines. The aim of the modelling is to generate a total model spectrum that reproduces the observed continuum (stellar plus nebular), the absolute emission line fluxes (with $I(\mathrm{H}\beta)$ as reference), and the line ratios.   

\subsubsection{Modelling of the optical spectrum} 
We use TLUSTY's stellar atmosphere grid \citep{2003ApJS..146..417L} (Log $Z=0.0$), parametrized with effective temperature $T_\mathrm{eff}$, luminosity ($L$) and gravity ($g$), to quantify the input radiation from the central star. As discussed earlier (Sec. \ref{sec:specfeatures}), the central star temperature is previously estimated as $T_\mathrm{Z}=33000$ K, and also we detect no He~{\sc ii} lines in our spectrum of MaC 2-1. Hence, we choose to vary $T_\mathrm{eff}$ below $55000$ K. Luminosity is kept as a free parameter. Gravity is arbitrarily kept at Log $g=4.5$ and left unchanged during the modelling as the model results do not sensitively depend on $g$. Hence, we cannot constrain $g$ from modelling. We finally estimate it using the mass and the radius of the central star (Sec. \ref{sec:bestmodelmac2-1}).  

The radial hydrogen density distribution used in our CLOUDY model is based on the observed HST image of MaC 2-1 (Fig. \ref{fig:denprofilemac2-1}(a)). Due to the unavailability of narrow band image mapping the H emission, we use the [O~{\sc iii}] image as an approximation. We construct a simplistic 3D model (Fig. \ref{fig:denprofilemac2-1}(b)) of the nebula taking the HST image as reference. For this, we use the 3D modelling code SHAPE. We construct a 3D mesh model using \textit{sphere} geometry, with a spherical inner shell surrounded by a spherical outer shell, to represent the morphology as observed in the image. Using \textit{density} modifier, we attribute density to these shells. The code solves radiation transfer to generate 2D synthetic image of the nebula. We match the modelled image with the observed image by adjusting the radii of the components. From the best-fitting model according to eye-estimation, the inner and outer radii (in terms of angular size) of the modelled nebula is found to be at $r_\mathrm{in}=0.27^{\prime\prime}$ and $r_\mathrm{out}=1.5^{\prime\prime}$, respectively. The interface of the inner and outer shells is at $r_\mathrm{inter}\sim0.45^{\prime\prime}$. Next, we scan the radial intensity of the observed image along eight different position angles and take average of the profiles. We consider the average profile as the observed radial intensity profile of the nebula. We adjust the radial density structure of the 3D model until we obtain best eye-estimated fit of the observed and modelled radial intensity profiles. Finally, we obtain a constant density for the inner shell seen in the image, where the density is maximum. The density decrease radially outwards following certain values to match the observed profile. The entire radial density structure of the nebula is shown in Fig. \ref{fig:denprofilemac2-1}(c)) in orange dotted line.

The hydrogen density ($n_\mathrm{H}$) in our model is specified using the derived radial density structure discussed above. As $n_\mathrm{H}$ in the inner shell is varied, the $n_\mathrm{H}$ of the whole nebula varies according to the derived radial density structure. We vary $n_\mathrm{H}$ of the inner shell in a range of $\sim$3000-6000 cm$^{-3}$, such that the radial average of $n_\mathrm{H}$ remains close to the $N_\mathrm{e}$ obtained from the nebular lines (Sec. \ref{sec:pynebmac2-1}).      
  
Distance ($d$) to the PN is introduced as a model parameter since we aim to match the absolute fluxes in the spectrum. Previous reporting of distance to MaC 2-1 include $d=13.06\pm4.58$ kpc by FPB16; $d=20.52$ kpc by SSV16. With reference to these, we vary distance in a range of 10 to 20 kpc. Accordingly, for each distance values, absolute values for the radii (in parsecs) are calculated to match the observed angular size of the nebula (specified by $r_\mathrm{in}$ and $r_\mathrm{out}$) as obtained from the 3D SHAPE model discussed above.       

The initial elemental abundances are set to the values calculated using the direct method (Sec. \ref{sec:pynebmac2-1}). The abundances are tuned to correctly reproduce the fluxes of emission lines when necessary. As we have no CELs of C to constrain C/H from modelling, we adopt the \citep{2010Ap&SS.328..179G} value. However, we find that proper value of C/H might be essential to reproduce correct $T_\mathrm{e}$ within the nebula (see discussion in Sec. \ref{sec:C/H}). Thus, we get an indirect estimation of C/H from modelling.  

\subsubsection{Modelling of the dust features}
After obtaining the best-fit model to reproduce the optical spectrum, we attempt to model the mid-IR spectrum by including dust in the model. We consider dust to exist in the entire nebula. We model the dust continuum and features using the grain code in CLOUDY. To match the dust continuum, we consider amorphous carbon (AC), graphite, and silicate grains in the model. We assume that the grains are spherical and follow the ISM type grain size distribution, $n(a)\propto{a^{-3.5}}$ \citep{1977ApJ...217..425M}. The sizes (in terms of radius) are in the range of $0.005-0.25$ $\mu$m. The AC grain opacity data are obtained by compiling the refractive index data (wavelength vs. $n-k$, where n and k are the real and imaginary part of the refractive index) from \citet{1991ApJ...377..526R}. For graphite and silicate, we use the compiled grain opacity data within the CLOUDY distribution.   

To model the weak SiC feature (11.2 $\mu$m), we consider ellipsoidal grain shapes. We observe that the feature could not be fitted with spherical shaped grains. The refractive indices are taken from \citet{1993ApJ...402..441L}. The grain opacities, for a particular size and axes ratio (a:b:c), within the required energy range are calculated using the formalism given by \citet{1983aslsp.book.....B}. For each shape, we calculate opacities for ten different sizes, in the range of $0.005-0.25$ $\mu$m, and assuming ISM type size distribution. 

To fit the 30 $\mu$m MgS feature, we adopt MgS refractive index data from \citet{1994ApJ...423L..71B} for Mg$_{0.9}$Fe$_{0.1}$S molecule, given in the range 10-500 $\mu$m. We extrapolate the data down to lower wavelengths ($\sim$0.001 $\mu$m), following the pattern observed for SiC grains, to be able to compile the data in CLOUDY, and obtain opacities. We find that the observed broad emission feature cannot be generated using spherical shaped grains alone. We use a mixture of spherical with ellipsoidal grain shapes with different axes ratios. For each shape, we assume ten sizes, in the range of $0.005-0.25$ $\mu$m, with ISM type size distribution. 

\subsubsection{Results from best-fitting photoionization model} \label{sec:bestmodelmac2-1}
We obtain a good match between the observed and modelled emission line fluxes and the continuum (Fig. \ref{fig:pimodelmac2-1}). The difference between the logarithmic fluxes of modelled and observed emission lines (both in the scale of H$\beta=100$) are plotted along the wavelength in Fig. \ref{fig:errormac2-1}. The $T_\mathrm{e}$ is well-reproduced in the model. The model results are summarised in Table \ref{tab:pimodelmac2-1}. In Table \ref{tab:obsvsmodmac2-1}, the observed and modelled fluxes for different emission lines are listed side-by-side for comparison. 

The best-fitting model of MaC 2-1 comprises of a central star of $T_\mathrm{eff}=49500$ K and $L=3900$ $L_{\sun}$. We estimate $d=16$ kpc for the PNe (see Sec. \ref{sec:distance} for more details). Accordingly, the $r_\mathrm{in}$ and $r_\mathrm{out}$ of the nebula is obtained as $0.021$ and $0.116$ pc,respectively. The $n_\mathrm{H}$ in the inner shell is estimated to be $4.0\times10^3$ $\mathrm{cm^{-3}}$. Also, from the modelling, we find the nebula to be matter-bound, i.e, the nebula is fully ionized up to the outer radius (see discussion in Sec. \ref{sec:ionizationfront&PDR}).

The model estimated abundances (Table \ref{tab:pimodelmac2-1}) in the best-fitting model are obtained by tuning up or down the values derived using direct method. Hence, the model abundances are to be taken as the final abundances obtained by us in this paper. While our model abundances other than N/H are comparable to those obtained previously by \citet{1996A&A...307..215C} and \citet{2017MNRAS.471.4648V}, our estimated N/H is higher than their values. The N/H in \citet{1996A&A...307..215C} might had been an underestimation due to the assumption of $\mathrm{N/O}=\mathrm{N^+/O^+}$, which is not proper for lower ionization PNe \citep{2014MNRAS.440..536D}. The lower N/H in \citet{2017MNRAS.471.4648V} is presumably due to the possible underestimation of the N/H quantity, as extra-mixing was not considered by the authors while modelling the red giant branch evolution.

From dust modelling, we find the dust temperature in the range of $51-210$ K considering all dust types. The dust-to-gas mass ratio is estimated as $2.27\times10^{-4}$. The dust temperatures obtained for different dust species and their composition are listed in Table \ref{tab:pimodelmac2-1}. We find that the SiC feature fits well with two shapes with axes ratios, 3:1:1 and 18:1:1. We find best-fitting of MgS feature using spherical grains (1:1:1) along with ellipsoidal grains of axes ratios, 3:1:1, 8:1:1, and 18:1:1.

\begin{figure}
\centering
\scalebox{0.95}[0.95]{\includegraphics{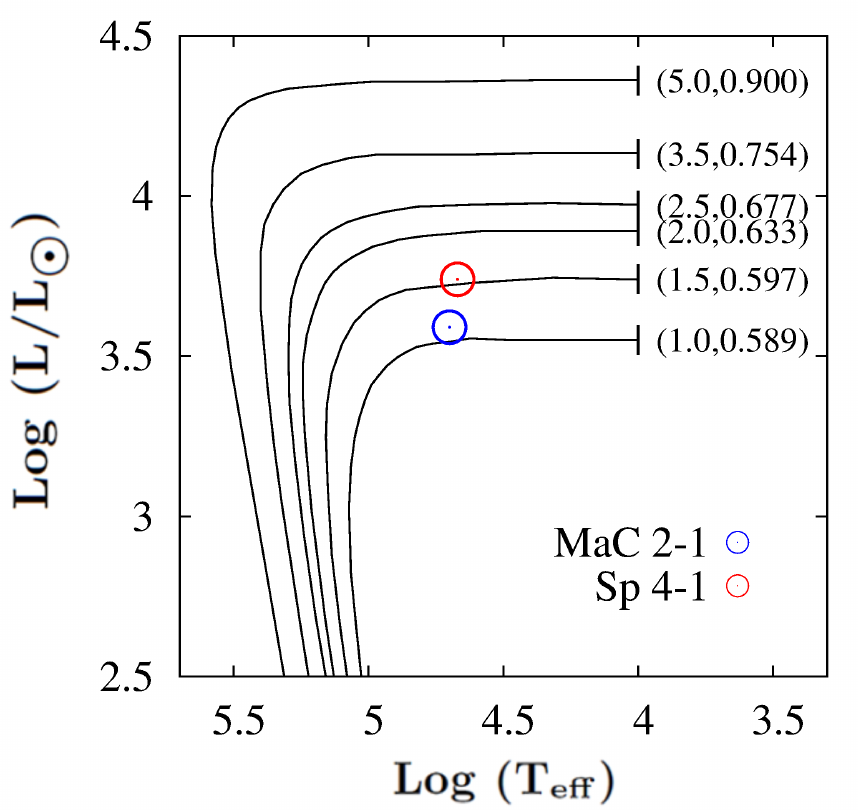}}

\scalebox{0.95}[0.95]{\includegraphics{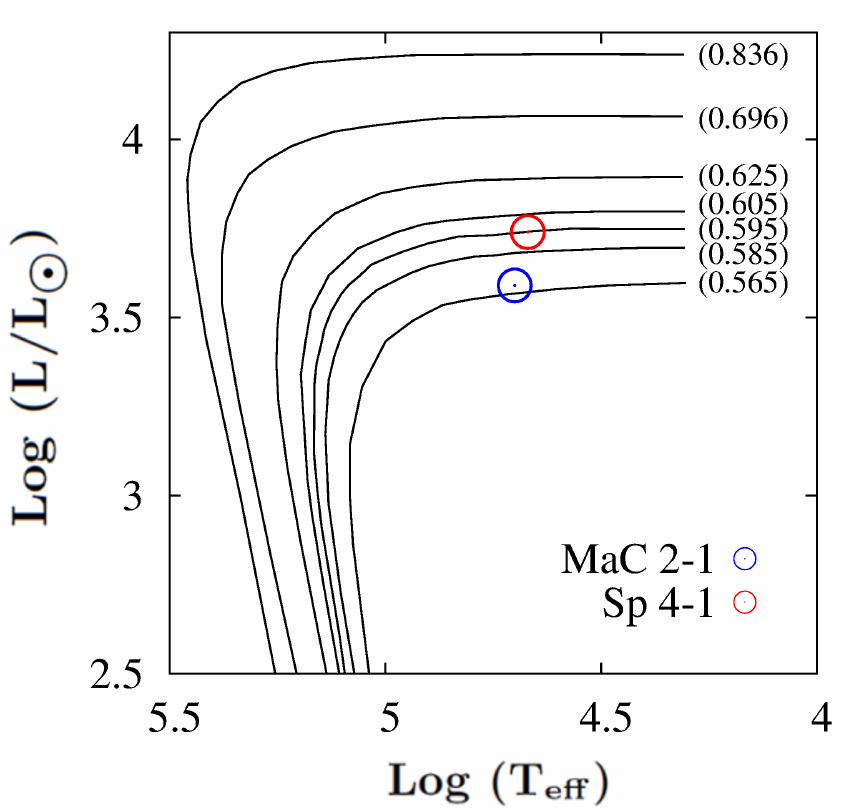}}
 \caption{Top: The model tracks (Log $(L/L_{\sun})$ vs. Log $(T_\mathrm{eff})$) shows the post-AGB phase of stellar evolutionary sequence, taken from \citet{1994ApJS...92..125V} ($Z=0.016$). Each track is labeled in the format (X, Y), where, `X' corresponds to the value of initial (main-sequence) mass and `Y' to the core mass at Log $(T_\mathrm{eff})=4$, in terms of solar masses. Bottom: The radiation-hydrodynamical evolutionary model tracks calculated by \citet{2013A&A...558A..78J}. Each track correspond to PN central star masses shown in brackets at the beginning of the tracks. In both plots, we place MaC 2-1 and Sp 4-1 among the model tracks at the co-ordinates (Log $(L/L_{\sun})$, Log $(T_\mathrm{eff})$) corresponding to our estimated values of $T_\mathrm{eff}$ and $L$ from photoionization modeling. Positions of the PNe are marked with circled points.}
 \label{fig:massfunction}
\end{figure}

We estimate the main-sequence mass (progenitor mass, $M_\mathrm{pr}$) of the central star from model tracks (Log $(L/L_{\sun})$ vs. Log $(T_\mathrm{eff})$) presenting post-AGB phases of stellar evolutionary sequences, computed by \citet{1994ApJS...92..125V} (Fig. \ref{fig:massfunction}). The range of models we use are for H burning PN Nuclei (PNN) of initial masses ranging from $1-5$ $M_{\sun}$, with $Z=0.016$. Using the values of $T_\mathrm{eff}$ and $L$ from our photoionization modeling, we estimate $M_\mathrm{pr}\sim1.2$ $M_{\sun}$ for MaC 2-1. To estimate the mass of the central star (remnant mass, $M$), we use radiation-hydrodynamical models by \citet{2013A&A...558A..78J} (Fig. \ref{fig:massfunction}), which are calculated for remnant masses of $0.565-0.836$ $M_{\sun}$ and thus, estimate $M\sim0.57$ $M_{\sun}$. From our $T_\mathrm{eff}$ and $L$, we further estimate the central star radius as, $R=0.85$ $R_{\sun}$, and consequently have Log $g=4.34$.

We use the method of determining quality factor ($\kappa_\mathrm{i}$) \citep{2009A&A...507.1517M} to test the quality of fitting of individual observables. The quality factor of the $i$th observable is defined as, 
\begin{equation}
\kappa_\mathrm{i}=\frac{\mathrm{Log}(M_\mathrm{i}/O_\mathrm{i})}{\tau_\mathrm{i}} \\ \& \\ \tau_\mathrm{i}=\mathrm{Log}(1+\sigma_\mathrm{i})
\end{equation}
where, $\sigma_\mathrm{i}$ denotes the error in the $i$th observable. We assume $\sigma_\mathrm{i}$ as $0.1$, if $I(\mathrm{\lambda})>0.1$ $\mathrm{H}\beta$; $0.2$, if $0.1$ $\mathrm{H}\beta>I(\mathrm{\lambda})>0.01$ $\mathrm{H}\beta$ and $0.3$, if $I(\mathrm{\lambda})<0.01$ $\mathrm{H}\beta$. These error values are defined in order to obtain $\kappa_\mathrm{i}$ corresponding to each emission line considering a tolerance limit of $\mid{\kappa_\mathrm{i}}\mid<1$. The quality factor corresponding to the emission lines of MaC 2-1 are given in Table \ref{tab:obsvsmodmac2-1}.

\section{Results: Study of Sp 4-1} \label{sec:sp4-1}
Sp 4-1 ($R.A.=19^h00^m26.67^s$, $Dec.=+38^{\circ}21^{\prime}04^{\prime\prime}.82$) appears as a round or circular shaped PN having a radius of about $0.6^{\prime\prime}$. It gets a classification of `round' PNe in SSV16 and reported to have `inner structure'. 

\subsection{Spectral analyses}

\begin{figure}
\centering
\scalebox{0.65}[0.65]{\includegraphics{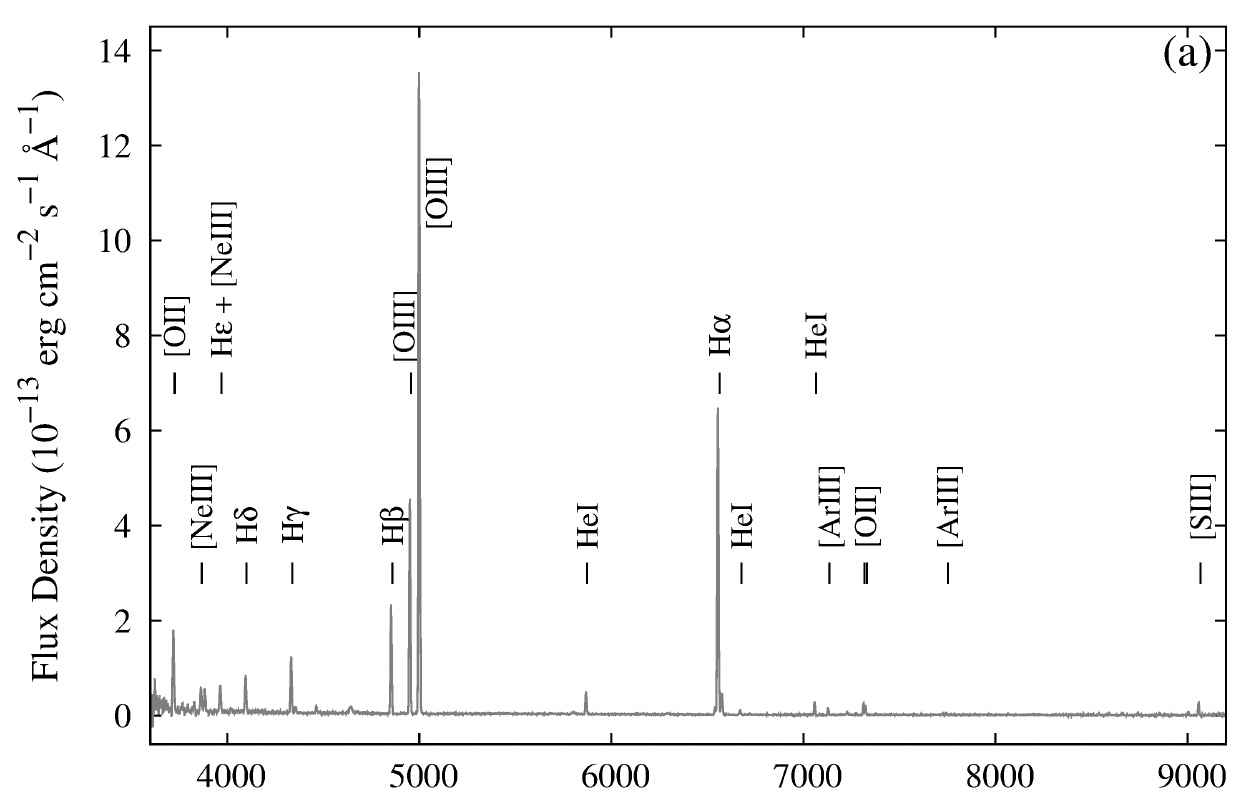}}

\scalebox{0.65}[0.65]{\includegraphics{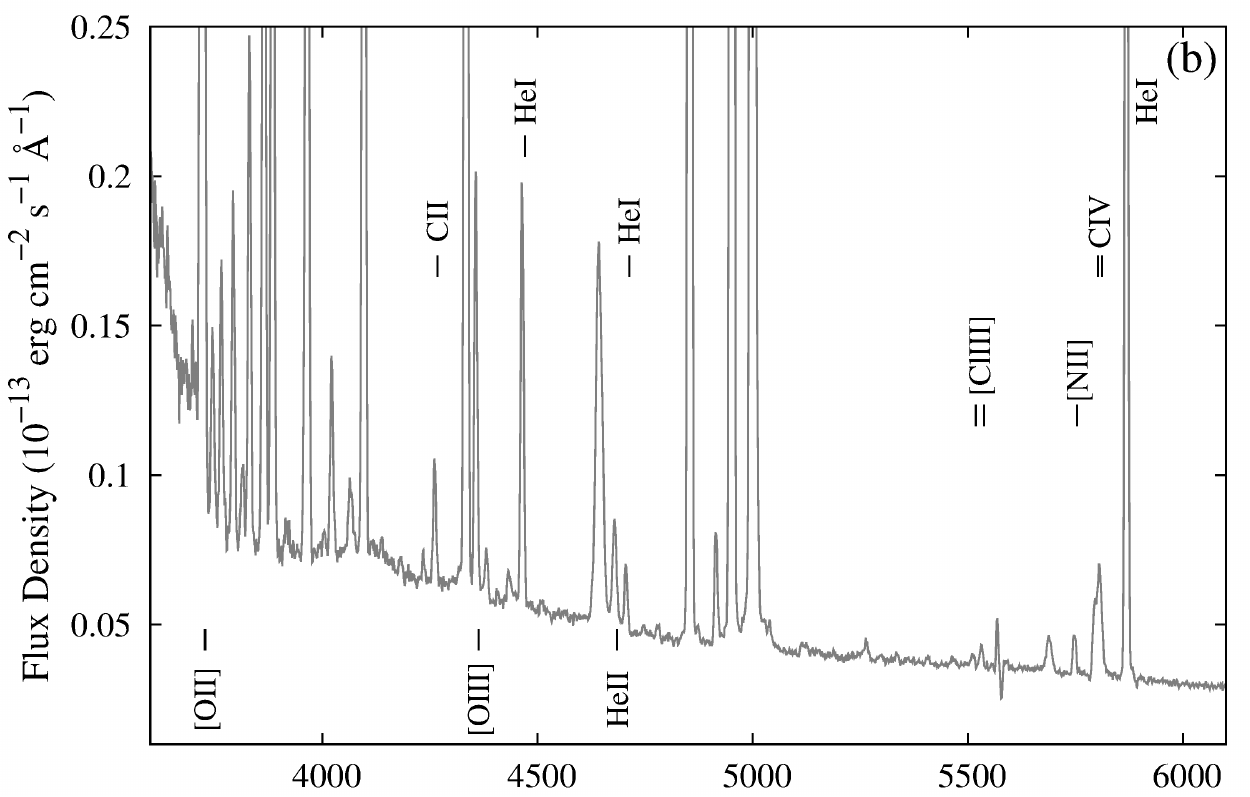}}

\scalebox{0.65}[0.65]{\includegraphics{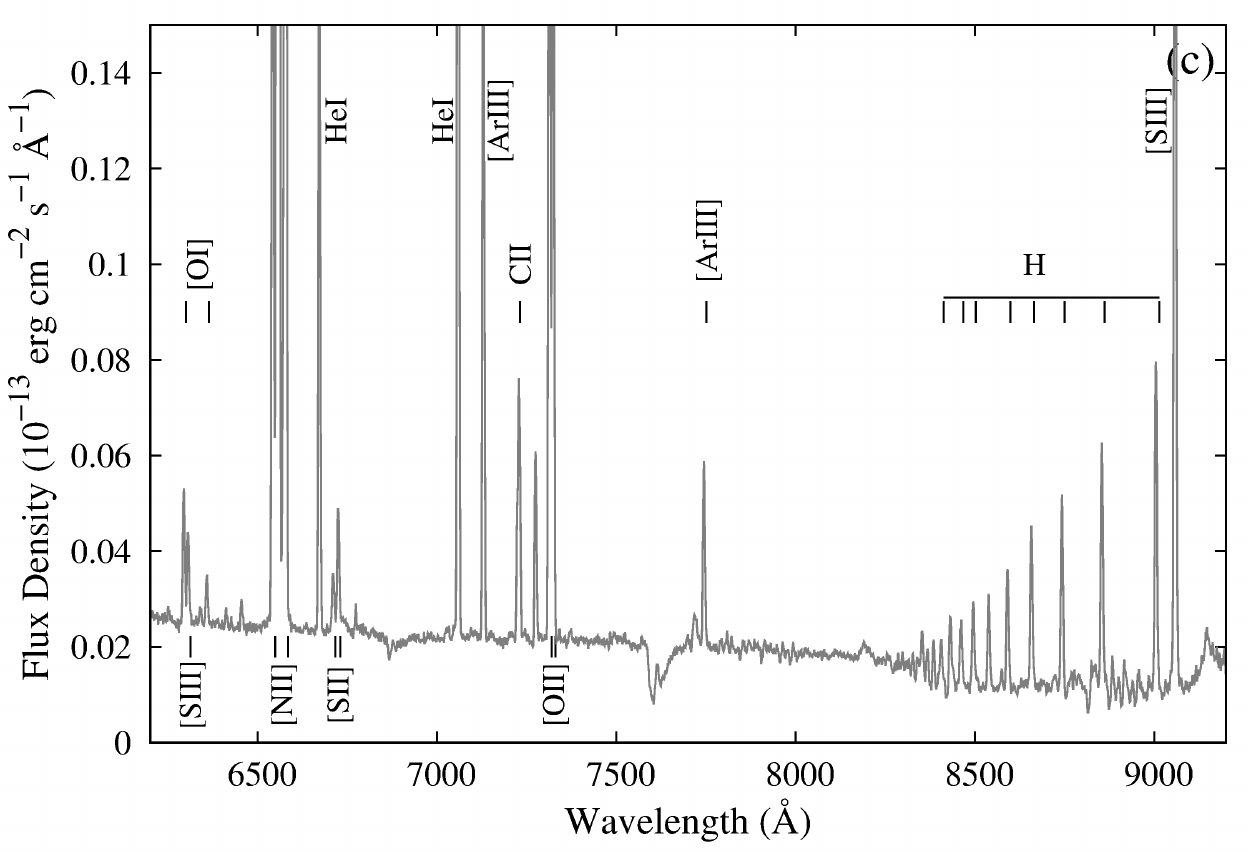}}
 \caption{The optical spectrum of Sp 4-1 in intrinsic absolute fluxes. Panel (a) shows the entire spectrum with strongest emission lines. The panels (b) and (c) show the magnified spectrum to emphasize the weaker emission lines in the blueward and redward regions, respectively. \label{fig:specoptsp4-1}}
\end{figure}

\subsubsection{Interstellar extinction coefficient}
Using Equation 1, we calculate the values, $c(\mathrm{H}\beta)_{\mathrm{H}\alpha}=0.18$, $c(\mathrm{H}\beta)_{\mathrm{H}\gamma}=0.11$, and from Equation 2, we have $c(\mathrm{H}\beta)_{\mathrm{H}\delta}=0.15$ and $c(\mathrm{H}\beta)\simeq0.17$. \citet{1992A&AS...95..337T} obtained $c(\mathrm{H}\beta)$ for Sp 4-1 as 0.39. While this is considerably higher than our value, \citet{2005MNRAS.362..424W} (hereafter W05) reported $c(\mathrm{H}\beta)=0.0$. Hence,  $c(\mathrm{H}\beta)$ shows a variation, which is possible if there is a relative shift of the central wavelengths of nebular H emission and stellar H absorption lines, i.e., relative motion of the central star and nebula. For example, such time variation of $c(\mathrm{H}\beta)$ is observed for SaSt 2-3 \citep{2019MNRAS.482.2354O}, where binarity of the central star has been suggested. However, for Sp 4-1, this can only be confirmed through time series observation of the central star in future. 

\subsubsection{Spectral features} \label{sec:specsp4-1}
{\textit{Optical spectrum}:} The overall optical spectrum (Fig. \ref{fig:specoptsp4-1}) of Sp 4-1 shows a low to moderately ionized PN. The usual RLs of H and He are strongly present in the spectrum. \citet{2016A&A...593A..29M} (hereafter M16) estimated Zanstra temperatures of central star as $37900\pm230$ K from H~{\sc i} flux and $59200\pm460$ K from He~{\sc ii} flux, which may imply the nebula to be optically thin for H ionizing radiation. However, in Sp 4-1 the He~{\sc ii} emission might not have originated in nebula (discussed below). M16 also estimated a high luminosity $\sim$ $13000^{+11000}_{-6000}$ $L_{\sun}$ for the central star. 

On the basis of observed spectral features in its optical spectrum, the central star of Sp 4-1 has been earlier classified as a weak emission line star (\textit{wels}) \citep{1993A&AS..102..595T,2011A&A...526A...6W}. In our observed spectrum as well, we observe features, which have known to be the characteristics of \textit{wels} \citep{2003AJ....126..887M}. For example, the 4650 {\AA} blend (N~{\sc iii}, C~{\sc ii}, and C~{\sc iv}), the C~{\sc iv} 5802, 5812 {\AA} doublet, and the C~{\sc iii} 5696 {\AA} emission are quite strong, similar to those observed in \textit{wels} PNe. We also detect few other characteristic lines found in \textit{wels} PNe, such as 5470 {\AA} blend (C~{\sc iv} and O~{\sc iv}) and the C~{\sc ii} 5892 {\AA} line. Presently, the \textit{wels} classification is not clear as \citet{2016MNRAS.458.2694B} showed in case of a few PNe (earlier known to have \textit{wels} central stars) that the lines characterizing a \textit{wels} are actually of nebular origin. In \citet{2018A&A...614A.135W} the classification of the central star of Sp 4-1 is updated as probable H rich O-type star. However, we clearly observe that the \textit{wels} lines show about twice the full width at half maximum (FWHM) of a nebular line in the spectrum of Sp 4-1. This is particularly evident from the prominently detected C~{\sc iii} 5696 {\AA} line, which is not a blend either. At this point, we may at least suggest that in case of Sp 4-1, the \textit{wels} lines are not of nebular, but of stellar origin, where the exact mechanism is not clear. 

This also leads to an uncertainty over the origin of the He~{\sc ii} 4686 {\AA} line detected in our spectrum. The nebular He~{\sc ii} 4686 {\AA} is a characteristics of moderate to high excitation class PNe. However, this line can also be formed in the stellar wind. One way of solving this uncertainty is to obtain accurate excitation class of the PN. Still, primary and most reliable excitation classification scheme depends on the intensity of nebular He~{\sc ii} 4686 {\AA} line. M84 proposed a secondary scheme for excitation class using the [Ne~{\sc iii}] 3869 line. From our spectrum of Sp 4-1, we have $I$([Ne~{\sc iii}] 5007)/$I(\mathrm{H}\beta)=0.18$. According to M84, this ratio should be $\leq0.25$ for the excitation classes $0-3$. Comparing this to A56 and D90 classifications, Sp 4-1 belongs to the excitation classes where no nebular He~{\sc ii} 4686 {\AA} should be detected. This may lead to the fact that the He~{\sc ii} 4686 {\AA} line present in our spectrum is of stellar wind origin. The prominent CELs observed in our spectrum of Sp 4-1 are, [N~{\sc ii}] 6548, 6583 {\AA} and the auroral [N~{\sc ii}] 5755 {\AA}; [O~{\sc iii}] 5007, 4959 {\AA} and the auroral [O~{\sc iii}] 4363 {\AA}; [O~{\sc i}] 6300, 6364 {\AA}; [Ne~{\sc iii}] 3869 {\AA}; [S~{\sc iii}] 6312, 9069 {\AA}; [Ar~{\sc iii}] 7136, 7751 {\AA}; and the doublets [S~{\sc ii}] 6716, 6731 {\AA}; [Cl~{\sc iii}] 5518, 5538 {\AA}, and [Ar~{\sc iv}] 4711, 4740 {\AA}.

\begin{figure}
\centering
\scalebox{0.65}[0.65]{\includegraphics{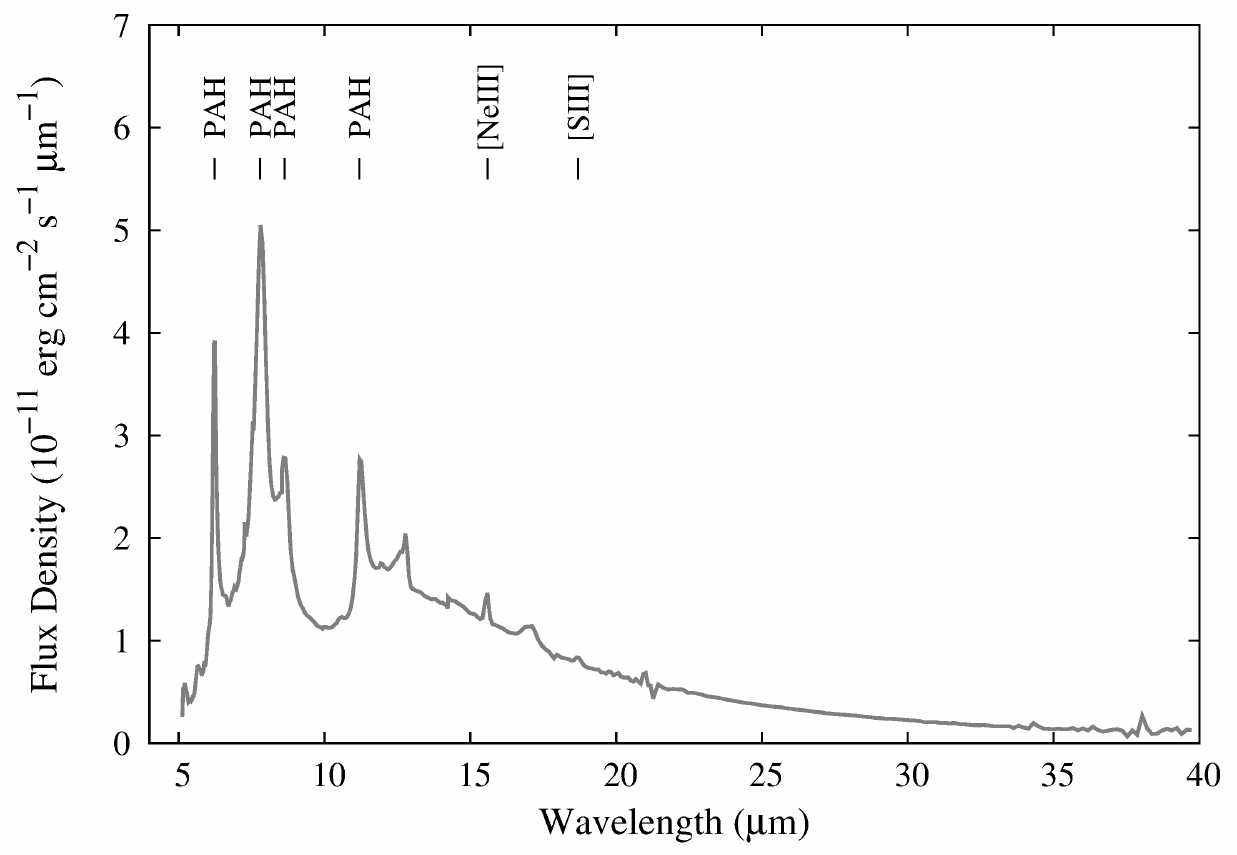}}
 \caption{The mid-IR Spitzer spectrum of Sp 4-1. Prominent features including the broad dust features are marked. Fluxes are in absolute scale. \label{fig:specmirsp4-1}}
\end{figure}
 
{\textit{IR spectrum}:} The shape of the mid-IR continuum in the spectrum (Fig. \ref{fig:specmirsp4-1}) suggests thermal dust emission from carbonaceous grains. As reported by S12, Sp 4-1 have carbon rich dust of aromatic nature. Strong and prominent 6.2, 7.7, 8.6, and 11.2 $\mu$m PAH emission features are crowned over the dust continuum. The precise peaks of the charged PAH emissions are at 6.23, 7.8, 8.63 $\mu$m, which makes Sp 4-1 to fall in the Class B according to the classification developed by \citet{2002A&A...390.1089P}, and \citet{2004ApJ...611..928V}. The fluxes of the PAH emissions are estimated using PAHFIT \citep{2007ApJ...656..770S}. [Ne~{\sc iii}] 15.6 $\mu$m and [S~{\sc iii}] 18.7 $\mu$m are detected but are weak. 

\begin{figure}
\scalebox{0.5}[0.5]{\includegraphics{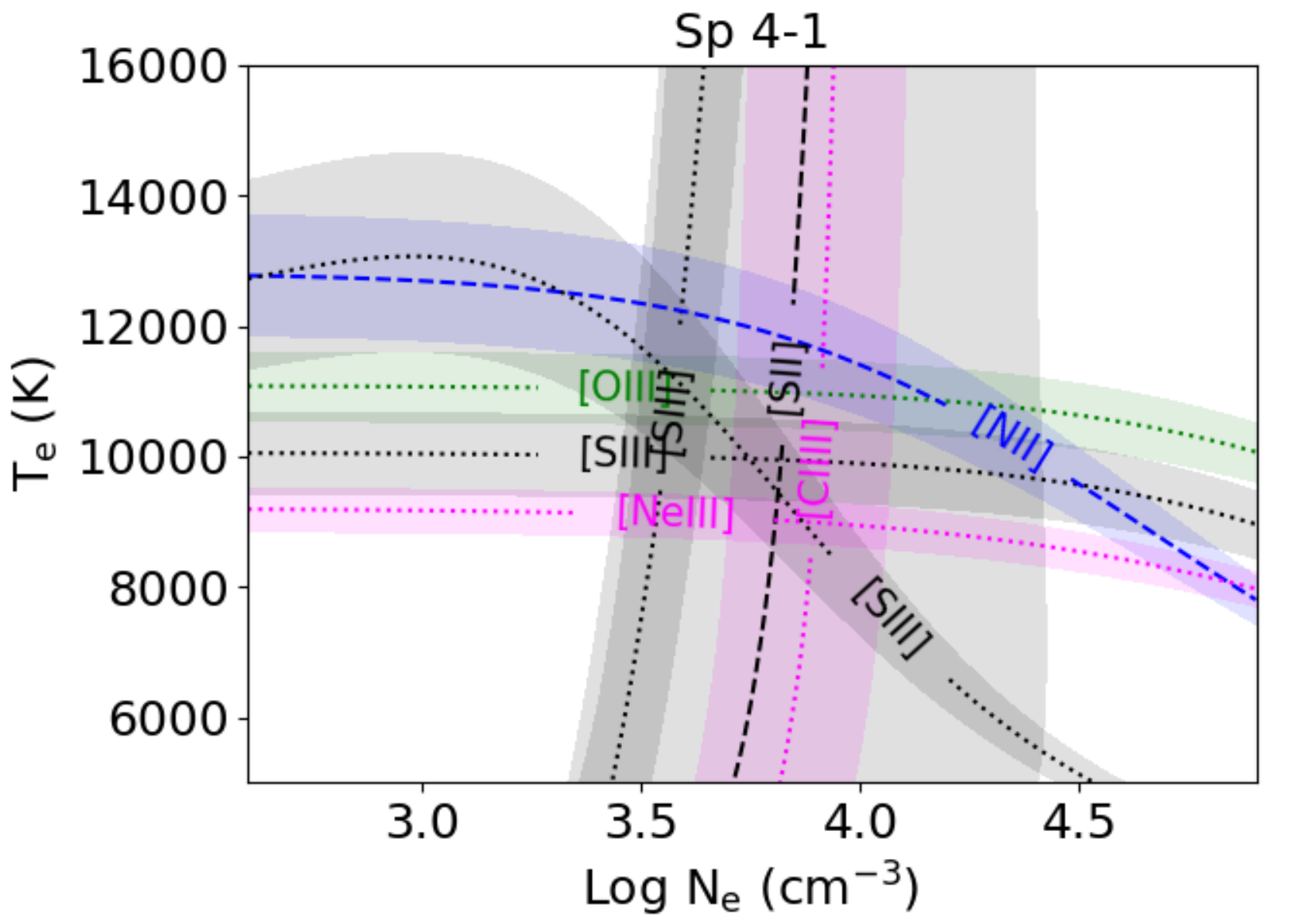}}
 \caption{The $T_\mathrm{e}$ vs Log $N_\mathrm{e}$ plot for Sp 4-1. The labels with the graphs denote the corresponding ion used for generating the same. \label{fig:tenesp4-1}}
\end{figure} 

\begin{table}
\centering
\small
\caption{Temperatures and Densities for Sp 4-1. \label{tab:tenesp4-1}}
 \begin{tabular}{lccc}
 \hline
& Diagnostic line id. & Ratio & $T_\mathrm{e}$/$N_\mathrm{e}$ value\\
 \hline
$T_\mathrm{e}$ & [O~{\sc iii}] 4363/5007 & 0.009 & ${10980}$\\
$T_\mathrm{e}$ & [S~{\sc iii}] 6312/9069 & 0.067 & ${9870}$\\
$T_\mathrm{e}$ & [N~{\sc ii}] 5755/6584 & 0.026 & ${11850}$\\
$T_\mathrm{e}$ & [Ne~{\sc iii}] 3869/15.6m & 0.429 & ${9030}$\\
$T_\mathrm{e}$ & [S~{\sc iii}] 9069/18.7m & 0.833 & ${9550}$\\
 \hline
$N_\mathrm{e}$ & [Cl~{\sc iii}] 5538/5518 & 1.472 & ${8550}$\\
$N_\mathrm{e}$ & [S~{\sc ii}] 6731/6716 & 1.917 & ${6590}$\\
 \hline
 \end{tabular}
\end{table} 

\begin{table}
\centering
\small
\caption{Nebular ionic and total abundances of Sp 4-1 from direct method ($A(B)=A\times10^B$). \label{tab:abunsp4-1}}
\begin{tabular}{l c c c c c}
\hline
X & X$^{+}$/H & X$^{2+}$/H & X$^{3+}$/H & ICF & X/H\\
\hline
He	&	0.115	&	-	&	-	& 1.0 	& 0.115   \\
C	&	-		&	-		&	-   & - 		& -       \\
N	&	4.03(-6)	&	-		&	-	& 10.5 	& 4.22(-5)\\
O	&	6.27(-5)	&	1.63(-4)	&	-	& 1.0 	& 2.26(-4)\\
Ne	&	-		&	1.45(-5)	&	-	& 2.65 	& 3.80(-5)\\
S	&	6.09(-8)	&	9.95(-7)&	-	& 1.07 		& 1.12(-6)\\
Cl	&	-		&	3.33(-8)	&	-	& 1.38 		& 4.55(-8)\\
Ar	&	-		&	3.51(-7)&	-	& 1.18 	& 4.11(-7)\\
\hline
\end{tabular}
\end{table}

\begin{table}
\centering
\small
\caption{Model results for Sp 4-1. \label{tab:pimodelsp4-1}}
\begin{tabular}{l c c c c r}
\hline
Geometry &&&&& Spherical\\
$d$ $(\mathrm{kpc})$ &&&&& 18.0\\
$T_\mathrm{eff}$ $(\mathrm{K})$ &&&&& 47000\\
$L$ $(L_{\sun})$ &&&&& 5500\\
$R$ $(R_{\sun})$ &&&&& 1.12\\
$M$ $(M_{\sun})$ &&&&& $\sim$0.6\\
$M_\mathrm{pr}$ $(M_{\sun})$ &&&&& 1.55\\
Log $g$ $(\mathrm{cm}$ $\mathrm{s^{-2}})$ &&&&& 4.12\\
$r_\mathrm{in}$ $(\mathrm{pc})$ &&&&& 0.026\\
$r_\mathrm{inter}$ $(\mathrm{pc})$ &&&&& 0.037\\
$r_\mathrm{ion}$ $(\mathrm{pc})$ &&&&& 0.057\\
$r_\mathrm{out}$ $(\mathrm{pc})$ &&&&& 0.07\\
$n_\mathrm{H}$ (at $r_\mathrm{in}$) $(10^4$ $\mathrm{cm^{-3}})$ &&&&& 1.15 \\
$f$ &&&&& 1.0\\
$m_\mathrm{cloud}$ $(M_{\sun})$ &&&&& 0.14\\
\hline
\hline
\end{tabular}
\begin{tabular}{l c c c c}
\multicolumn{5}{c}{Chemical composition}\\
\hline
& \multicolumn{3}{c}{Model ionic fractions} &\\
X & X$^{+}$ & X$^{2+}$ & X$^{3+}$ & X/H \\
\hline
He	&	0.9931	&	0.0000	&	-	&	0.12	\\
C	&	0.0875	&	0.9078	&	0.0042		&	1.35(-3)	\\
N	&	0.1368	&	0.8551	&	0.0059		&	2.57(-5)\\
O	&	0.1820	&	0.8091	&	0.0000		&	1.67(-4)	\\
Ne	&	0.2153	&	0.7834	&	0.0000		&	1.87(-5)\\
S	&	0.0589	&	0.8433	&	0.0982		&	1.32(-6)	\\
Cl	&	0.0630	&	0.8690	&	0.0673		&	4.58(-8)\\
Ar	&	0.0164	&	0.9141	&	0.0692		&	3.64(-7)	\\
\end{tabular}
\begin{tabular}{l c c c c}
\hline
Dust types & $T_\mathrm{d}$ (K) & $(D/G)$ & X & (X/H)$_{dust}$\\
\hline
AC & 104-210 & 5.59(-4) & C & 6.98(-5)\\
PAHc & 164-270 & 2.15(-3) &  & \\
PAHn & 172-338 & 9.55(-4) &  & \\
All dust & 104-338 & 3.66(-3) & & \\
\hline
\end{tabular}
\end{table}

\begin{table}
\centering
\small
\caption{Comparison between observed and modeled line fluxes of Sp 4-1. Observed and modelled flux values are given with respect to $I(\mathrm{H}\beta)=100$. \label{tab:obsvsmodsp4-1}}
\begin{tabular}{lccc}
\hline
Observables & Observed & Modelled & $\kappa_\mathrm{i}$\\
& flux & flux &\\
\hline
Log $I(\mathrm{H}\beta)$ & -11.71 & -11.54 & \\
	H$\delta$			4101	{\AA}	&	26.44	&	26.34	&	0.04	\\
	H$\gamma$			4340	{\AA}	&	47.65	&	47.38	&	0.06	\\
	H$\beta$			4861	{\AA}	&	100.00	&	100.00	&	0.00	\\
	H$\alpha$			6563	{\AA}	&	290.52	&	277.41	&	0.48	\\
$\mathrm{	He~{\sc	I}	}$	4026	{\AA}	&	2.55	&	2.91	&	-0.71	\\
$\mathrm{	He~{\sc	I}	}$	4388	{\AA}	&	0.60	&	0.76	&	-0.89	\\
$\mathrm{	He~{\sc	I}	}$	4471	{\AA}	&	5.54	&	6.30	&	-0.70	\\
$\mathrm{	He~{\sc	I}	}$	5876	{\AA}	&	18.28	&	18.88	&	-0.34	\\
$\mathrm{	He~{\sc	I}	}$	6678	{\AA}	&	4.51	&	4.61	&	-0.12	\\
$\mathrm{	He~{\sc	I}	}$	7065	{\AA}	&	9.04	&	14.76	&	-2.69	\\
$\mathrm{	He~{\sc	I}	}$	7281	{\AA}	&	1.32	&	1.25	&	0.30	\\
$\mathrm{	[N~{\sc	II}]	}$	5755	{\AA}	&	0.58	&	0.59	&	-0.02	\\
$\mathrm{	[N~{\sc	II}]	}$	6548	{\AA}	&	4.43	&	7.49	&	-2.88	\\
$\mathrm{	[N~{\sc	II}]	}$	6583	{\AA}	&	21.71	&	22.08	&	-0.18	\\
$\mathrm{	[O~{\sc	I}]	}$	6300	{\AA}	&	1.10	&	1.19	&	-0.44	\\
$\mathrm{	[O~{\sc	I}]	}$	6364	{\AA}	&	0.35	&	0.38	&	-0.32	\\
$\mathrm{	[O~{\sc	II}]	}$	3727	{\AA}	&	59.32	&	57.78	&	0.28	\\
$\mathrm{	[O~{\sc	II}]	}$	7325	{\AA}	&	14.60	&	11.49	&	2.52	\\
$\mathrm{	[O~{\sc	III}]	}$	4363	{\AA}	&	5.46	&	6.45	&	-0.91	\\
$\mathrm{	[O~{\sc	III}]	}$	4959	{\AA}	&	205.73	&	204.10	&	0.08	\\
$\mathrm{	[O~{\sc	III}]	}$	5007	{\AA}	&	607.32	&	608.93	&	-0.03	\\
$\mathrm{	[Ne~{\sc	III}]	}$	3869	{\AA}	&	18.12	&	18.92	&	-0.45	\\
$\mathrm{	[Ne~{\sc	III}]	}$	15.50	$\mu$m	&	64.53	&	16.84	&	14.09	\\
$\mathrm{	[Ne~{\sc	III}]	}$	36.00	$\mu$m	&	10.45	&	1.37	&	21.33	\\
$\mathrm{	[S~{\sc	II}]	}$	6716	{\AA}	&	0.41	&	0.49	&	-0.66	\\
$\mathrm{	[S~{\sc	II}]	}$	6731	{\AA}	&	0.79	&	0.80	&	-0.03	\\
$\mathrm{	[S~{\sc	III}]	}$	6312	{\AA}	&	0.64	&	0.81	&	-0.91	\\
$\mathrm{	[S~{\sc	III}]	}$	9069	{\AA}	&	9.44	&	9.64	&	-0.11	\\
$\mathrm{	[S~{\sc	III}]	}$	18.70	$\mu$m	&	17.33	&	4.62	&	13.88	\\
$\mathrm{	[S~{\sc	III}]	}$	33.50	$\mu$m	&	5.86	&	0.80	&	10.93	\\
$\mathrm{	[Cl~{\sc	III}]	}$	5518	{\AA}	&	0.23	&	0.25	&	-0.38	\\
$\mathrm{	[Cl~{\sc	III}]	}$	5538	{\AA}	&	0.33	&	0.52	&	-1.68	\\
$\mathrm{	[Ar~{\sc	III}]	}$	7136	{\AA}	&	5.08	&	5.06	&	0.02	\\
$\mathrm{	[Ar~{\sc	III}]	}$	7751	{\AA}	&	1.29	&	1.20	&	0.37	\\
	PAH			6.2	$\mu$m	&	797.67	&	793.96	&	0.05	\\
	PAH			7.7	$\mu$m	&	2754.74	&	2393.01	&	1.48	\\
	PAH			8.6	$\mu$m	&	783.76	&	497.15	&	4.78	\\
	PAH			11.2	$\mu$m	&	697.5	&	503.65	&	3.42	\\
\hline
\end{tabular}
\end{table}

\subsubsection{Electron temperatures, densities and CEL anbundances} \label{sec:pynebsp4-1}
Using PyNeb we obtain plasma properties, $T_\mathrm{e}$ and $N_\mathrm{e}$, and the ionic and total elemental abundances within the nebula. $T_\mathrm{e}$ are calculated using [O~{\sc iii}], [S~{\sc iii}], [Ne~{\sc iii}], and [N~{\sc ii}]. $N_\mathrm{e}$ are calculated using [Cl~{\sc iii}] and [S~{\sc ii}]. Table \ref{tab:tenesp4-1} lists the line ratios used for the calculation of $T_\mathrm{e}$ and $N_\mathrm{e}$ with their calculated values. From the $T_\mathrm{e}$-$N_\mathrm{e}$ plot (Fig. \ref{fig:tenesp4-1}), we approximate the values, $T_e\sim10800$ K and $N_e\sim7000$ cm$^{-3}$ for our further calculation of abundances using ICFs \citep{2014MNRAS.440..536D}. The ionic abundances of the elements and the total elemental abundances obtained with direct method are listed in Table \ref{tab:abunsp4-1}.

\begin{figure}
\centering
\scalebox{0.4}[0.4]{\includegraphics{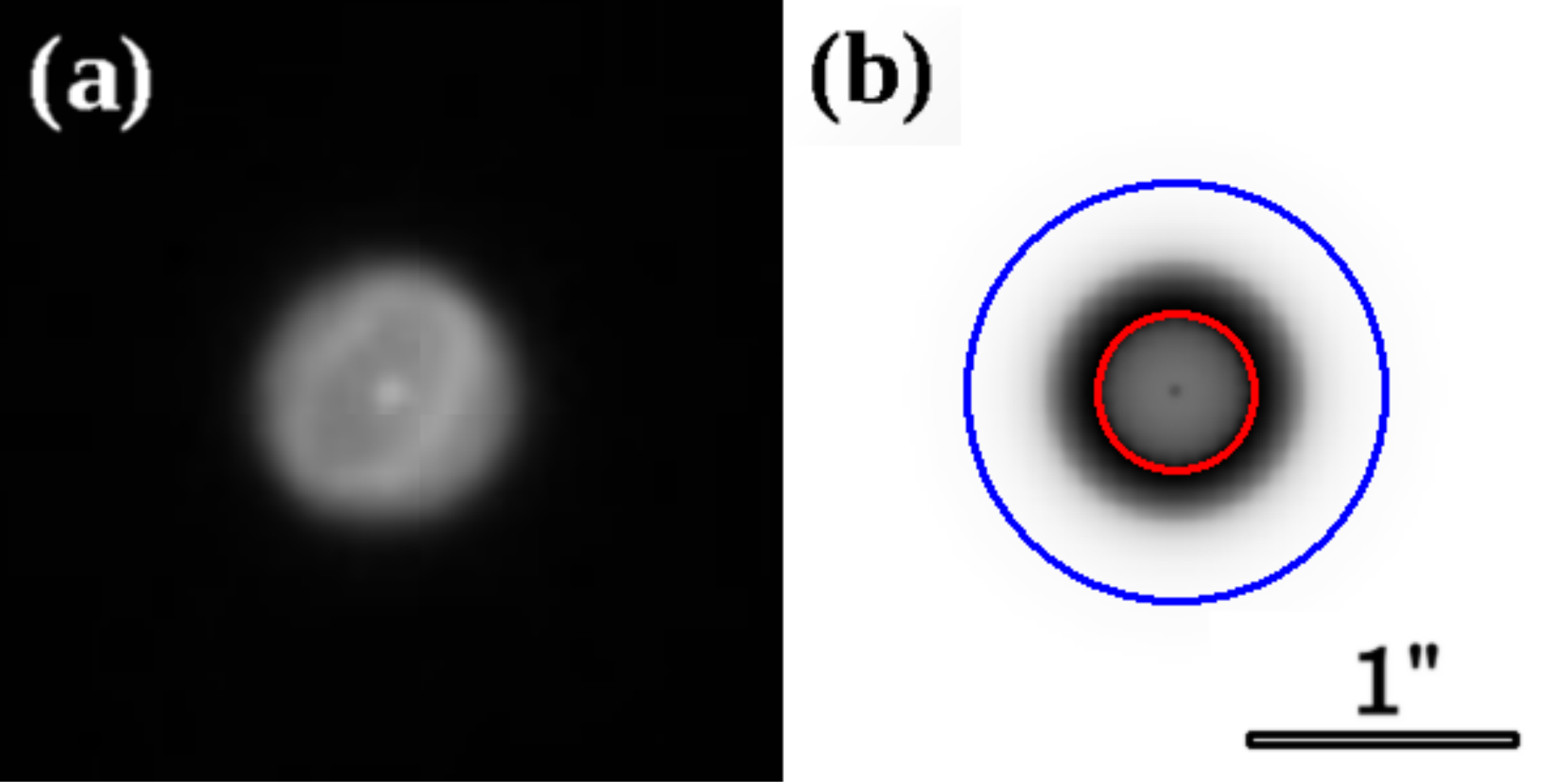}}

\scalebox{0.9}[0.9]{\includegraphics{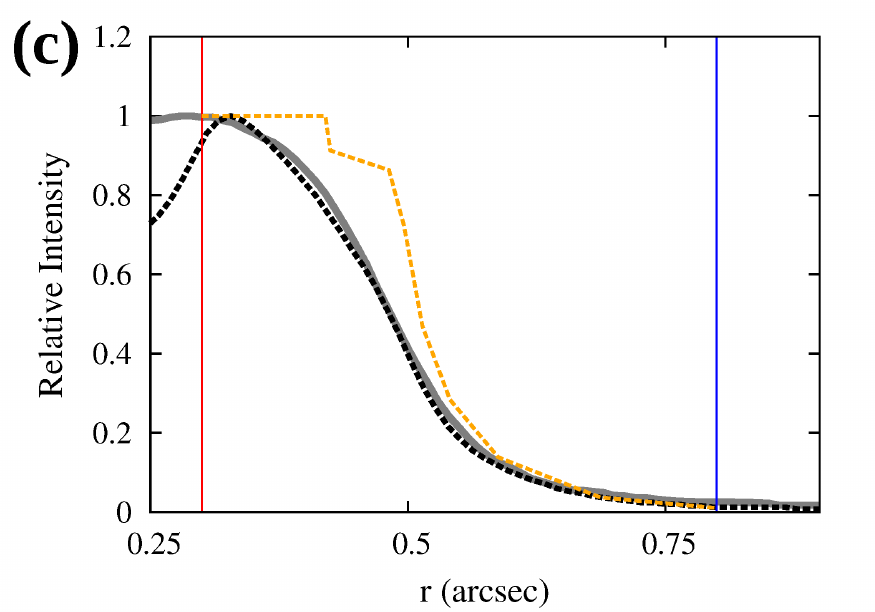}}
 \caption{(a) The HST image of Sp 4-1 in [O~{\sc iii}] from SSV16. The image box measures $3^{\prime\prime}\times3^{\prime\prime}$. (b) The grey scale 3D model image of Sp 4-1 constructed in SHAPE. The concentric circles are the inner (in red) and outer (in blue) radii of the shell. (c) The grey solid line presents the observed radial intensity profile obtined from the HST image. The orange dotted line shows the radial density structure of the 3D model, that best-fits the observed radial intensity profile. The best-fitting modelled radial intensity profile (black dotted line) shows a good fit with the observed profile. The vertical red and blue lines denote the inner and outer radii, respectively, as mentioned above. In our CLOUDY modelling, the density of the shell is specified using the radial density struture obtained from the 3D model. Also, the $r_\mathrm{in}$ and $r_\mathrm{out}$ are calculated using the inner and outer radii of the 3D model (see details in Sec. \ref{sec:pimodelsp4-1}). \label{fig:denprofilesp4-1}}
\end{figure} 

\begin{figure*}
\centering
\scalebox{1.0}[1.0]{\includegraphics{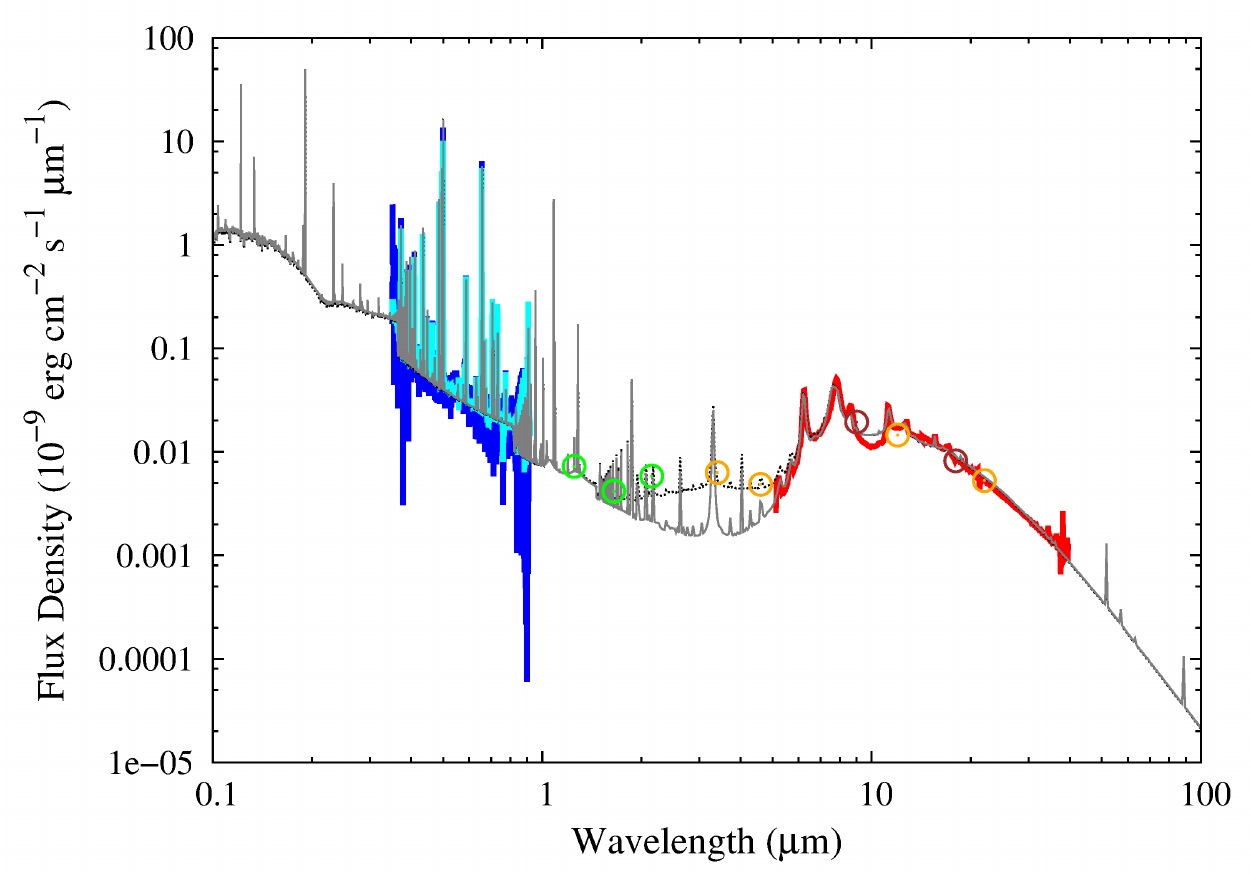}}
 \caption{The modelled spectra of Sp 4-1 in grey solid line is plotted with the observed data for comparison. The observed optical spectrum is shown in blue (210 s spectrum) and cyan (2100 s spectrum), and mid-IR spectrum is shown in red. The green, orange and brown circles denote the observed photometric data points from 2MASS, WISE and AKARI catalogues, respectively. The black dotted line is the test model spectrum that includes warm dust region (see details in Sec. \ref{sec:testmodels}) \label{fig:pimodelsp4-1}}
\end{figure*}

\begin{figure}
\centering
\scalebox{0.63}[0.63]{\includegraphics{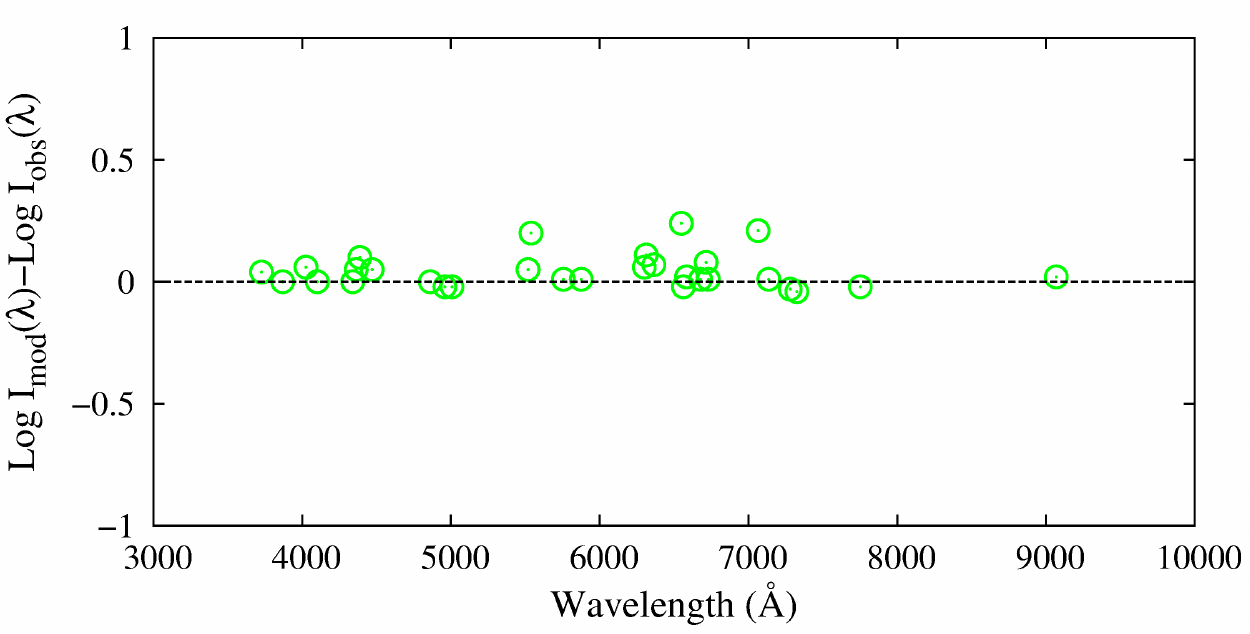}}
 \caption{The differences between logarithmic observed and modelled emission line flux are plotted along the wavelength. The points falling close to the zeroth line implies a good model fit. \label{fig:errorsp4-1}}
\end{figure}  

\subsection{Modelling of Sp 4-1} \label{sec:pimodelsp4-1} 
We compute a photoionization model of Sp 4-1 using CLOUDY to fit the observables and estimate the parameters involving the system self-consistently. We aim to match the model generated spectrum with the observed optical spectrum and mid-IR \textit{Spitzer} spectrum of Sp 4-1. The basic modelling assumptions and procedure is the same as described for the modelling of MaC 2-1 (Sec. \ref{sec:pimodelmac2-1}).         

\subsubsection{Modelling of the optical spectrum}
In Sec. \ref{sec:specsp4-1} we see that $T_\mathrm{eff}$ might be much lower than that estimated from He~{\sc ii} by M16, as He~{\sc ii} might not have generated within nebula. We varied $T_\mathrm{eff}$ up to an upper limit of 65000 K. For the model atmosphere, we use TLUSTY up to $T_\mathrm{eff}=55000$ and `Rauch PN Nuclei' \citep{2003A&A...403..709R} for larger $T_\mathrm{eff}$s (both with Log $Z=0.0$). $L$ is varied as a free parameter. We arbitrarily choose Log $g=4.5$ during the modelling and estimate it using the mass and the radius of the central star as obtained for MaC 2-1.

We notice that we cannot generate the observed He~{\sc ii} flux in the model, while trying to match other observables. Recalling the possible wind origin of He~{\sc ii} in the spectrum, we do not attempt further to fit the He~{\sc ii} flux in the model. 

It is to be noted from the image of Sp 4-1 (Fig. \ref{fig:denprofilesp4-1}(a)) that it seems to have a very faint halo around its prominently observed structure. In the SHAPE modelling, we find that in order to fit the observed radial intensity profile up to the base of the profile, it is required that nebula should extent to a larger $r_\mathrm{out}$ than the prominent boundary ($\sim0.6^{\prime\prime}$) observed in the HST image. In Fig. \ref{fig:denprofilesp4-1}(b), we show the modelled image of the nebula, obtained from 3D modelling, in greyscale. The best-fitting 3D model gives the radii as $r_\mathrm{in}=0.3^{\prime\prime}$, $r_\mathrm{inter}\sim0.42^{\prime\prime}$ and $r_\mathrm{out}=0.8^{\prime\prime}$. The observed and radial intensity profile, fitted with the modelled radial intensity profile, is shown in Fig. \ref{fig:denprofilesp4-1}(c)). The radial density structure of the modelled nebula is shown in Fig. \ref{fig:denprofilesp4-1}(c)) in orange dotted line. The inner shell have a constant radial density, followed by a sharp decline at the interface, then, a nearly plateau region (representing the rarer outer shell). Beyond that, the density falls sharply and the nebula becomes very faint. In the CLOUDY model, we vary $n_\mathrm{H}$ (for inner-shell) in a range of $\sim$5000-20000 cm$^{-3}$, i.e., in close vicinity of $N_\mathrm{e}$ obtained from the nebular lines (Sec. \ref{sec:pynebsp4-1}). 

The only reported distance to Sp 4-1 we find is 21.51 kpc by SSV16. With reference to this value, we vary $d$ within the range $\sim$15-25 kpc in our models.           

\subsubsection{Modelling of the dust features}
We consider that dust is distributed in the entire nebula. To match the dust continuum, we assumed spherical AC grains that follow the ISM type grain size distribution. The size range, i.e., the minimum and maximum sizes of the grains are varied. 

We compile the opacity data from \citet{2007ApJ...657..810D} to model the PAH emission features seen in the spectrum. The abundances of charged and neutral PAH molecules (denoted by PAHc and PAHn, respectively) are varied to match the relative strengths of observed charged and neutral features. The sizes of the PAH grains range from $4.3\times10^{-4}$ (30 C atoms) to $1.1\times10^{-3}$ $\mu$m (500 C atoms), with an ISM type size distribution.      

\subsubsection{Results from best-fitting photoionization model} 
Our modelled spectrum of Sp 4-1 agrees well with the observed spectrum (Fig. \ref{fig:pimodelsp4-1}). Overall, the plasma diagnostic line ratios are well-reproduced in the model. The differences between logarithmic line fluxes (both in the scale of H$\beta=100$) are depicted in Fig. \ref{fig:errorsp4-1}. We summarise the model results in Table \ref{tab:pimodelsp4-1}. The observed and the modelled fluxes of different emission lines are listed in Table \ref{tab:obsvsmodsp4-1} for comparison, along with the $\kappa_\mathrm{i}$ values. 

From the best-fitting model of Sp 4-1, we estimate the parameters of the central star as, $T_\mathrm{eff}=47000$ K and $L=5500$ $L_{\sun}$, and from these, we have $R=1.12$ $R_{\sun}$. We estimate a distance of $d=18$ kpc for the PN (see Sec. \ref{sec:distance} for more details). Hence, $r_\mathrm{in}$ is obtained as $0.026$ pc. We find the nebula to be fully ionized and hence radiation-bound. The ionization front is at $r_\mathrm{ion}=0.057$ pc according to the best-fitting model, which closely matches observed boundary of the nebula in the HST image (Fig. \ref{fig:denprofilesp4-1}(a)). Hence, the nebula seems to have a photodissociation region (PDR). In our CLOUDY models, we run our calculations up to $r_\mathrm{out}$ value of $0.07$ pc, which is the outer boundary according to the 3D model for the estimated distance. The $n_\mathrm{H}$ in the inner shell is estimated to be $1.15\times10^4$ $\mathrm{cm^{-3}}$.

We compare our model estimated abundances with the abundances reported by W05. Our estimated abundances are consistent for N/H, O/H, Ar/H and S/H with their estimated values. Our estimated Ne/H is about half of that obtained by W05. 

We find that the mid-IR continuum of Sp 4-1 best fits with AC dust size of $0.001-0.007$ $\mu$m. We obtain dust temperature varying within $104-338$ K considering all types of dust. The dust-to-gas mass ratio is estimated as $3.66\times10^{-3}$. The detailed dust composition of the nebula are listed in Table \ref{tab:pimodelsp4-1}.  

Using the similar method as described for MaC 2-1, we estimate the progenitor mass, $M_\mathrm{pr}\sim1.55$ $M_{\sun}$ (Fig. \ref{fig:massfunction}) using the model tracks from \citet{1994ApJS...92..125V} (Fig. \ref{fig:massfunction}), and our model values of $T_\mathrm{eff}$ K and $L$. We estimate the mass of the central star, $M\sim0.6$ $M_{\sun}$ using the models from \citet{2013A&A...558A..78J} (Fig. \ref{fig:massfunction}), and hence, we obtain Log $g=4.12$ for the central star.

\section{Discussion} \label{sec:discussion}

\subsection{Probable degeneracies in the models}
As photoionization models use a large number of parameters, degeneracy may arise among certain free parameters. Independent estimation of one or more of these parameters may solve the degeneracies. Also, fitting of enough number of observables with a large number of models reduces the chance of degenerate solution. In our modelling, we aim to appropriately reproduce the absolute observed fluxes (and flux ratios), and hence, the overall ionization of the nebula. This, in turn depends on appropriate temperatures and densities. Also, we aim to fit the angular size of the nebula, which relates size and distance, aided with an well-estimated density function. With these observables as references, we run a large number of models with different sets of parameters, and interpret them properly in physical terms and thus, attempt to constrain $T_\mathrm{eff}$, $L_{\sun}$, $d$ (and $r$) and $n_\mathrm{H}$ in our models. Further degeneracy may arise among $L$, $d$ (and $r$) and $f$, if $f$ is a free model parameter \citep{2009A&A...507.1517M}. To avoid this degeneracy, we do not consider clumping and keep the value of $f$ at 1.0 throughout the modelling.

\subsection{Accuracy in distance to the PNe} \label{sec:distance}
As discussed, we have assumed ``no-clumps" ($f=1.0$) scenario for the nebulae to reduce uncertainty in $d$. However, in a realistic model, we may need to consider a finite amount of clumping. To test the validity of ``no-clumps" assumption, we apply the independent method of distance estimation used in \citet{2009A&A...507.1517M}, which solves the possible degeneracy among $L$, $d$ and $f$. 

The said parameters in two degenerate models (1 and 2) are theoretically related as, ${d_2}=x{d_1}$, ${L_2}=x^2{L_1}$ and ${f_2}={f_1}/x$ ($x$ is a real number). We also say that the model 2 parameters are the true values and same as obtained by the independent method. From \citet{1994ApJS...92..125V}, using the values computed for `H-burning PNN evolutionary models' ($Z=0.016$), a relation between the nebular age ($t$) and $L$ can be obtained at model estimated value of $T_\mathrm{eff}$ as,  
\begin{equation}
\mathrm{Log}(t_2)=K_1-K_2\mathrm{Log}(L_2)
\end{equation}
As the degeneracy in $d$ extends to $t$ (considering $r={v_\mathrm{exp}}t$, where $v_\mathrm{exp}$ is nebular expansion velocity), we have ${t_2}=x{t_1}$. Thus, 
\begin{equation}
\mathrm{Log}(t_2)=K_3+0.5\mathrm{Log}(L_2) 
\end{equation}
Here, $K_3=\mathrm{Log}(t_1/0.5{L_1})$.                                                                                                                                                                                                                                                                                                                                                                                                                                                                                                                                                                                                                                                                                                                                                                                                                                                                                                                                                                                                                                                                                                                                                                                                                                                                                                                                                                                                                                                                                                                                                                                                                                                                                                                                Solving equations 3 and 4, we can obtain ${L_2}$ and determine $x$.

To obtain $v_\mathrm{exp}$, we use the high-resolution spectra (in the range of $6540-6590$ {\AA}) of the PNe received through private communication with Prof. J. A. L{\'o}pez (see \citet{SPMCatalog}). From the FWHM of the H$\alpha$ line profiles, we estimate $v_\mathrm{exp}$ values as $\sim$17 and $\sim$20 km s$^{-1}$ for MaC 2-1 and Sp 4-1, respectively. We use the $r_\mathrm{out}$ values of the nebula obtained from our modelling to calculate $t$. For MaC 2-1, we obtain $x=1.05$, using $r_\mathrm{out}=0.116$ pc.                                                                                                                                                                                                                                                                                                                                                                                                                                                                                                                                                                                                                                                                                                                                                                                                                                                                                                                                                                                                                                                                                                                                                                                                                                                                                                                                                                                                                                                                                                                                                                                                                                                                                                                                                                                                                                                                                                                                                                                                                                                                                                                                                                                                                                                                                                                                                                                                                                                                                                                                                                                                                                                                                                                                                                                                                                                                                                                                                                                                                                                                                                                                                                                                                                                                                                                                                                           
For Sp 4-1, using $r_\mathrm{out}=0.07$ pc, we get $x=0.997$.                                                                                                                                                                                                                                                                                                                                                                                                                                                                                                                                                                                                                                                                                                                                                                                                                                                                                                                                                                                                                                                                                                                                                                                                                                                                                                                                                                                                                                                                                                                                                                                                                                                                                                                                                                                                                                                                                                                                                                                                                                                                                                                                                                                                                                                                                                                                                                                                                                                                                                                                                                                                                                                                                                                                                                                                                                                                                                                                                                                                                                                                                                                                                                                                                                                                                                                                                                                                                                                                                                                                                                                                                                                                                                                             
Hence, the filling factors are indeed close to unity, which indicates the validity of the model assumption $f=1.0$, and hence, validates the estimation of $d$ from modelling.

\subsection{Indirect estimation of C/H from modelling} \label{sec:C/H}
As mentioned before, C/H is not estimated from direct method due to the absence of CELs of C in the spectral range we use. For the same reason, we cannot directly constrain C/H from modelling. However, we get an indirect estimation of C/H from modelling of the optical spectrum for both the PNe. We notice that without choosing a proper C/H in the model, we could not properly reproduce the ionization balance inside the nebulae, particularly in case of the ionization balance of O, while reproducing the [O~{\sc iii}] 4363/5007 ratio. It is important to mention that we use tuning of abundances only as secondary option to reproduce the correct ionization structure, after finding the best solution for other model parameters ($T_\mathrm{eff}$, $L_{\sun}$, etc.). However, error in this abundance estimation might be high and hence, should only be considered to get a qualitative idea. For example, from the modelling, we obtain $\mathrm{C/O}=0.33$ for MaC 2-1 and $\mathrm{C/O}=8.6$ for Sp 4-1, where, both values are at the extreme ends comparing to the values found in literature. Considering high amount of error might have involved in these estimations, we only conclude, $\mathrm{C/O}<1$ for MaC 2-1 and $\mathrm{C/O}>1$ for Sp 4-1, as a result of the modelling of the optical spectrum. Note that C/H (and C/O) estimated this way should correspond to the CELs, as $T_\mathrm{e}$ is involved. Future observation of ultraviolet CELs of C may provide conclusive results on the CEL abundance of C for these PNe. Also, while concluding about the total $\mathrm{C/O}$ in the nebular environment, may need to consider the amount depleted in the dust, or the RL calculated abundances as well (see Sec. \ref{sec:dustchemistry}).
 
\subsection{Dust chemistry and C/O} \label{sec:dustchemistry}
Our modelling estimates carbon rich dust chemistry for both MaC 2-1 and Sp 4-1, as earlier reported by S12. It is expected that carbon rich dust chemistry in PNe would correspond to $\mathrm{C/O}>1$, while oxygen rich dust chemistry would correspond to $\mathrm{C/O}<1$ (e.g., \citealt{2007ApJ...671.1669S}). Exceptions are observed in case of galactic PNe, as mentioned below, which are attributed to error in the estimation of C/O from nebular lines, or inability to get the actual C/O from optical spectra due to depletion of C and O into dust (e.g., \citealt{2014ApJ...784..173D}; \citealt{2015MNRAS.449.1797D}). Here, MaC 2-1 shows such exception, as we estimate carbon rich dust chemistry and $\mathrm{C/O}<1$ for this PN. Examples of well-known PNe having carbon rich dust chemistry but also showing $\mathrm{C/O}<1$ from CEL calculations are NGC 6826 and NGC 3242 \citep{2014ApJ...784..173D,2017MNRAS.471.4648V}. However, $\mathrm{C/O}$ quantities calculated using RLs may differ from their CEL values (see Table 5. in \citealt{2014ApJ...784..173D}). For example, in case of NGC 3242, RL calculation gives opposite prediction: $\mathrm{C/O}>1$. MaC 2-1 may show similar results, which, at the moment is out of the scope of this paper due to the unavailability of required RLs of O in our spectrum. Assuming the possible scenario of different origin of RLs within the nebula, the problem may be addressed in the future for MaC 2-1 (and similar PNe), for example, using photoionization models with multiple regions within the nebula.

\subsection{Ionization front and PDR} \label{sec:ionizationfront&PDR}
We inspect the extent of ionized region and presence of PDR around the ionized region in the PNe. In case of MaC 2-1, we run test model with our best fit parameters without specifying $r_\mathrm{out}$, and stopping the model at 4000 K. We find that the ionization front could extend approximately to twice the $r_\mathrm{out}$ obtained from the model or the boundary observed in the image. We could not find a model for MaC 2-1 that fits the outer boundary and also match the line fluxes properly. Thus, MaC 2-1 seems to be matter-bound, i.e., fully ionized nebula without having a PDR region. This is in line with the fact that the neutral [O~{\sc i}] lines are not detected in our optical spectrum of MaC 2-1. Future inspection of neutral IR lines such as [O~{\sc i}] 63,146 $\mu$m and [C~{\sc ii}] 158 $\mu$m may give more evidence on the fact. On the other hand, we find that Sp 4-1 might be an radiation-bound nebula, as our best-fitting photoionzation model extends beyond ionization front. However, the H$\beta$ flux in our model of Sp 4-1 is higher than its observed value by $\sim$35$\%$. We could not lower the flux in the models to fit the observed value, while maintaining the otherwise well-fitted ionization structure. We may attribute the lower observed flux to certain deviation from spherical shape, while the model is spherical. For example, the nebula might actually be compressed pole-on or have open lobes in observer's direction.  

\subsection{Presence of warm dust within the nebulae: Test models} \label{sec:testmodels}
In some PNe, a sudden rise in the near-IR continuum close to $\sim$ $1.5-5$ $\mu$m is seen, which is attributed to warmer dust region within the nebula. For both PNe in this paper, we observe such characteristic from the position of photometric data points. The characteristic is more prominent in case of Sp 4-1 than in MaC 2-1. However, this can be confirmed by near-IR spectra of these PNe. Still, we attempt to obtain test models to include this characteristic in the computed models of these PNe. For both the PNe, we consider a narrow (Log ${\Delta}r\sim15.3$) and dense ($n_\mathrm{H}\sim10^5$ $\mathrm{cm^{-3}}$) region with AC dust of 0.1 $\mu$m size close to the central star. The dust-to-gas mass ratio in the warmer dust region is two orders higher in case of Sp 4-1 ($D/G=1.6\times10^{-3}$) than in MaC 2-1 ($D/G=2.3\times10^{-5}$). Dust temperatures (radial average) are obtained as 685 K for Sp 4-1 and 532 K for MaC 2-1 in the warmer dust region. In Figures \ref{fig:pimodelmac2-1} and \ref{fig:pimodelsp4-1}, we show the spectrum (black dotted line) obtained with the warmer dust region, for MaC 2-1 and Sp 4-1, respectively. We should mention that the inclusion of the warm dust region does not influence our previous results obtained without considering this region. 

\subsection{Peimbert class of the PNe}
PNe are classified in Types I-IV through a main classification scheme proposed by \citet{1978IAUS...76..215P}. According to galactic population, PNe are distributed in disk, bulge and halo. From this work, we find a metal-poor chemical composition for both the PNe. The abundances of O, Ne, S, Cl, and Ar, estimated from our modelling are much lower than the averages found in cases of galactic disk and bulge PNe \citep{2007MNRAS.381..669W}. The lower metallic abundances are similar to those found in type-I PNe \citep{1987RMxAA..14..540P}. However, in type-I PNe, the He and N abundances are found to be much higher than those estimated by us for MaC 2-1 and Sp 4-1. Also, our O/H values of MaC 2-1 and Sp 4-1 are not as low as for Type-IV (halo population) PNe, for which the limiting condition is suggested as $\mathrm{Log(O/H)}+12<8.1$ (e.g., \citealt{2003PASP..115...67O}). We further estimate the peculiar velocities ($V_\mathrm{pec}$) of the PNe from the equation given by \citet{2013RMxAA..49...87P}. We take the galactocentric distances ($R_\mathrm{G}$) reported in SSV16. However, our estimated distances are lower than those given by SSV16. Hence, we approximately scale down the value of $R_\mathrm{G}$ corresponding to our distances. We refer the heliocentric velocities ($V_\mathrm{hel}$) from \citet{SPMCatalog}. These give us $V_\mathrm{pec}\sim250$ km s$^{-1}$ for MaC 2-1 and $V_\mathrm{pec}\sim-200$ km s$^{-1}$ for Sp 4-1. Hence, both MaC 2-1 and Sp 4-1 seem to be Type-III PNe, having $V_\mathrm{pec}$ much higher than the limiting value of $60$ km s$^{-1}$ considered for this class. 

\section{Summary and conclusion} \label{sec:summary}
In this work, we study two of the very less-studied PNe, MaC 2-1 and Sp 4-1. Both the PNe falls in the category of low- to moderate-excitation PNe. Sp 4-1 is supposed to be have a emission line central star which might be a \textit{wels} type. Along with strong dust continuum emission, MaC 2-1 depicts SiC and MgS features and Sp 4-1 shows PAH bands in their mid-IR spectrum. We calculate the plasma properties ($T_\mathrm{e}$ and $N_\mathrm{e}$), ionic abundances and total elemental abundances from the emission line flux analyses. We obtain a compact description of the ionization structure of the PNe through photoionization modelling using CLOUDY. The models are well-constrained by simultaneous agreement of a sufficiently high number of characteristic observables with their modelled values. From modelling, we estimate temperature, luminosity and gravity of the ionizing central star. We evaluate nebular dimensions and its density structure. The chemical composition, in terms of total elemental abundances and dust composition are obtained. We estimate distances 16 and 18 kpc for MaC 2-1 and Sp 4-1, respectively. The progenitor masses are obtained as $1.2$ $M_{\sun}$ for MaC 2-1 and $1.55$ $M_{\sun}$ for Sp 4-1. We find both the PN to be metal-poor referring to the findings in literature. From morphological perspective, the PN images are well-reproduced by considering spherical shapes, and the spectrum are well-matched using spherical morphological considerations. However, a detailed kinematical information about nebula would be necessary for confident evaluation of 3D nebular structure and resolve degeneracies related to the actual dimension along the line of sight. Our estimated abundances are consistent with other findings. Also, the goodness of the models in reproducing a significant number of observables may imply that we have obtained reliable chemical composition for the PNe.           
 
\section*{Acknowledgements}
The authors are thankful to the anonymous reviewer for valuable comments and suggestions. We acknowledge S. N. Bose National Centre for Basic Sciences under Department of Science and Technology (DST), Govt. of India, for providing necessary support to conduct research work. We are thankful to the HCT Time Allocation Committee (HTAC) for allocating nights for observation, and the supporting staff of the observatory. We are grateful to Prof. Peter van Hoof for helpful discussion regarding the grain code in CLOUDY. We are thankful to Prof. J. A. L{\'o}pez for kindly providing the requested spectra. This work uses data based on observations made with the Spitzer Space Telescope, which is operated by the Jet Propulsion Laboratory, California Institute of Technology under a contract with NASA. This paper uses data based on observations made with the NASA/ESA Hubble Space Telescope, and obtained from the Hubble Legacy Archive, which is a collaboration between the Space Telescope Science Institute (STScI/NASA), the Space Telescope European Coordinating Facility (ST-ECF/ESA) and the Canadian Astronomy Data Centre (CADC/NRC/CSA). We thank the people maintaining these data bases.

\section*{Data Availability}
The optical spectra used in this paper will be shared on reasonable request to the corresponding author. The Spitzer data are available at the Spitzer Heritage Archive (SHA; https://sha.ipac.caltech.edu/). The HST data are available at the Hubble Legacy Archive (HLA; https://hla.stsci.edu/). The photometric data are available at the Infrared Science Archive (IRSA; https://irsa.ipac.caltech.edu).   

\bibliographystyle{mnras}
\bibliography{References}

\bsp	
\label{lastpage}
\end{document}